\documentclass[11pt,a4paper]{article}
\DeclareUnicodeCharacter{2223}{\mid}
\usepackage[english]{babel}
\usepackage{xcolor}
\usepackage[letterpaper,top=2cm,bottom=2cm,left=3cm,right=3cm,marginparwidth=1.75cm]{geometry}
\usepackage{array} %for table formatting
\usepackage{placeins} % for \FloatBarrier
\usepackage{float}
\usepackage{amsmath}
\usepackage{bm}
\usepackage[colorlinks=true, allcolors=blue]{hyperref}
\usepackage{graphicx}
\usepackage{gensymb}   %package to include the symbols
\usepackage{authblk}
\usepackage{units} % SI unit typesetting
\usepackage{url} % URL handling
\usepackage{longtable} % Tables that continue onto multiple pages
\usepackage{mathrsfs} % Support for \mathscr script
\usepackage{multirow} % Span rows in tables
\usepackage{multicol} % Span columns in tables
\usepackage{booktabs} % Enhanced tables with \toprule, etc.
\usepackage{makecell}
\usepackage{titlesec} % enable table of contents

\usepackage[utf8]{inputenc} % allow utf-8 input
\usepackage[T1]{fontenc}    % use 8-bit T1 fonts
\usepackage{hyperref}       % hyperlinks
\usepackage{fixmath}

\usepackage{booktabs}       % professional-quality tables
\usepackage{amsfonts}       % blackboard math symbols
\usepackage{nicefrac}       % compact symbols for 1/2, etc.
\usepackage{microtype}      % microtypography
\usepackage{lipsum}
\usepackage{lineno}         %package for using line numbers
\newcommand{\mat}[1]{\mathbf{#1}}

\title{Composition Design of Shape Memory Ceramics based on Gaussian Processes}
\author[a]{Ashutosh Pandey}
\author[b]{Justin Jetter}
\author[c]{Hanlin Gu}
\author[b]{Eckhart Quandt}
\author[a]{Richard James \footnote{Corresponding author: james@umn.edu}}

\affil[a]{Department of Aerospace Engineering and Mechanics, University of Minnesota, Minneapolis, MN, USA}
\affil[b]{Institute for Materials Science, Faculty of Engineering, Kiel University, Kiel, Germany}
\affil[c]{Department of Mechanics and Engineering Science, Peking University, Beijing, China}

\begin{document}
%\linenumbers %start the numbering
\date{}
\maketitle
\begin{abstract}

We present a Gaussian process machine learning model to accurately predict the transformation temperature and lattice parameters of ZrO$_2$-based ceramics doped with HfO$_2$, Y$_{0.5}$Ta$_{0.5}$O$_2$, and Er$_2$O$_3$.
Our overall goal is to search for a shape memory ceramic material with a reversible transformation and low hysteresis.
The input space of the model consists of key physical features derived from the electronic and bonding properties of the cations present in the compositions. 
The identification of a new low hysteresis composition is based on design criteria that have been successful in metal alloys: (1) $\lambda_2 = 1$, where $\lambda_2$
is the middle eigenvalue of the transformation stretch tensor \cite{cui_combinatorial_2006, zhang_energy_2009, zarnetta_identification_2010}, (2) minimizing the max$|q(f)|$, which measures the deviation from satisfying the cofactor conditions \cite{james_way_2005, Chen2013, gu2021exploding}, (3) high transformation temperature, (4) low transformational volume change, and (5) solid solubility. 
To identify a new composition satisfying the design criteria, we develop an algorithm to generate many synthetic compositions within the mole fraction ranges of the compositional space used to generate the training data for the model.
Using the predicted lattice parameters of the synthetic compositions, a strong correlation between $\lambda_2$ and max$|q(f)|$ is found. 
 We identify a promising composition, 31.75ZrO$_2$--37.75HfO$_2$--14.5Y$_{0.5}$Ta$_{0.5}$O$_2$--1.5Er$_2$O$_3$, which closely satisfies all the design criteria. However, differential thermal analysis reveals a relatively high thermal hysteresis of 137°C for this composition, indicating that the proposed design criteria are not universally applicable to all ZrO$_2$-based ceramics.
We also explore reducing tetragonality of the austenite phase by addition of Er$_2$O$_3$. The idea is to tune the lattice parameters of austenite phase towards a cubic structure will increase the number of martensite variants, thus, allowing more flexibility for them to accommodate high strain during transformation compared to tetragonal to monoclinic transformation. 
We find the effect of Er$_2$O$_3$ on tetragonality  is weak due to limited solubility. We conclude that a more effective dopant is needed to achieve significant tetragonality reduction.
Overall, Gaussian process machine learning models are shown to be highly useful for prediction of compositions and lattice parameters, but the discovery of low hysteresis ceramic materials apparently involves other factors not relevant to phase transformations in metals.\\

%Keywords: Phase Transformation, Shape Memory Ceramic, Cofactor Conditions, Gaussian Process, Machine Learning
\end{abstract}

\newpage
\tableofcontents

%define custom commands
\renewcommand{\vec}[1]{\mathbold{#1}}

\section{Introduction} \label{section1_introduction}
The shape memory effect is based on a thermally induced, reversible  phase transformation between a high-temperature austenite phase and a low-temperature martensite phase.
During a forward transformation, the austenite phase is cooled to a certain temperature
when the martensite first appears at the martensite start temperature $M_\text{s}$.
With further cooling, the austenite phase completely transforms into the martensite at the martensite finish temperature $M_\text{f}$.
The reverse transformation upon heating causes transformation back to the austenite, as
indicated by the austenite start $A_\text{s}$ and the austenite finish $A_\text{f}$ temperatures.
The austenite phase has a higher symmetry than the martensite phase, leading to symmetry-related variants of martensite \cite{Kaushik_2003}.
Due to compatibility with the austenite phase, the martensite phase typically appears as finely twinned laminates \cite{jian_electron_1995, jian_prediction_1997, pitteri_generic_1998} consisting of two variants of martensite with volume fractions $f$ and $(1-f)$.\\

Generally, a stressed transition layer separates the twinned martensite and austenite, leading to average compatibility between phases.
The reversibility of the phase transformation, measured by the width of the hysteresis loop \cite{zarnetta_identification_2010, knupfer_nucleation_2013} or by the
number of cycles to failure \cite{chluba_shape_2015, ni_exceptional_2016}, can be achieved by improving the compatibility of the two phases.
Generally, there is an equipartition between the elastic energy of the stressed transition layer and the total interfacial energy of the twin boundaries \cite{kohn_branching_1992, ball_fine_1987}.
The process of tuning geometric compatibility entails devising strategies to reduce the elastic energy in the transition layer by adjusting the lattice parameters via compositional changes \cite{delville_transmission_2010}. This is thought to improve the reversibility of the transformation by two distinct mechanisms: 1) remove an energy barrier associated with the creation of these stressed transition layers during transformation \cite{zhang_energy_2009} and 2) mitigate
 against plastic deformation in these
stressed layers as they move through
the material during transformation.\\
%%%%%%%%%%%%%%%%%%%%%%
Necessary and sufficient conditions for eliminating this transition layer in a simple austenite/martensite interface are $\lambda_2 = 1$ plus additional conditions, explained below.  Under these stronger conditions, 
stressed transition layers are not only eliminated for simple austenite/twinned martensite interfaces, but also in a wide variety of complex microstructures involving
multiple austenite/martensite interfaces
relevant to nucleation and growth. These
stronger conditions are termed conditions of 
 supercompatibility.
Within the theory of martensite \cite{Mackenzie_crystallography_1954, bowles_crystallography_1954} the {\it cofactor conditions}, introduced in \cite{james_way_2005, Chen2013} are notable conditions of supercompatibility.  Like $\lambda_2 = 1$ they are purely geometrical; they only involve the lattice parameters of the two phases.
The strategy of tuning lattice parameters by compositional changes has been used successfully in metals to lower thermal hysteresis to a few degrees K 
\cite{zarnetta_identification_2010, song_enhanced_2013, meng_thermal_2015} or to increase the 
fatigue life under repeated stress-induced
transformation (in tension) \cite{chluba_shape_2015} to
$10^7$ cycles.\\

Discovering a ceramic material with a phase transformation that could exhibit a reversible shape memory effect could enable diverse applications in aerospace engineering, biomedicine, and energy science.  
This discovery could serve as a basis for a reversible, low-hysteresis actuator that could function in high-temperature or corrosive environments. Since ceramics can exhibit ferroelectricity, this could potentially extend the set of known ferroelectrics, with interesting applications to energy conversion
in the small temperature difference regime \cite{wegner_correlation_2020}.
Among ceramics, $\text{ZrO}_2$-based ceramics are of special technological interest because the martensitic phase transformation in these materials is accompanied by a transformation strain of up to 10\% in shear and up to 3-5\% in volume. They undergo phase transformation from the austenite with a tetragonal crystal structure to the martensite with a monoclinic crystal structure. The achievable actuation stress in these ceramics is around $2 \times 10^{3} $ MPa \cite{lai_shape_2013}, offering high work output approaching 100 MJ/m$^3$, thus making it ideal for solid-state actuators.\\

Recently, there have been efforts in applying the cofactor conditions to Zirconia-based ceramics \cite{gu2021exploding, pop-ghe_suppression_2019, liang_tuning_2020, Pang2022} to search for a low thermal hysteresis shape memory ceramic (SMC).
The thermal hysteresis ($\mathrm{\Delta T}$) is determined as half the difference between the ($A_\text{s} + A_\text{f}$) and the ($M_\text{s} + M_\text{f}$).
Gu et al. \cite{gu2021exploding} discovered the lowest hysteresis in ceramic $(\text{Zr}_{0.45}\text{Hf}_{0.55}\text{O}_2)_{0.775}-(\text{Y}_{0.5}\text{Nb}_{0.5}\text{O}_2)_{0.225}$ with thermal hysteresis $\mathrm{\Delta T} = 134$ \degree C by employing new criteria based on minimizing the maximum deviation from the exact satisfaction of the cofactor condition; while Pang et al. \cite{Pang2022} proposed design criteria to guide the discovery of the shape memory ceramics with low hysteresis, which are ``(1) commensurate interfaces between transforming phases (close satisfaction of cofactor condition) (2) low transformation volume change (3) solid solubility (4) high transformation temperature ($M_\text{s}$ > 500 $\degree \text{C}$).''
By assessing how different dopants influence the Zirconia based SMC, they identified specific compositions that satisfied the design criteria and reduced thermal hysteresis.
They discovered that Zirconia doped with 17Ti--3Al--6Cr and with 20Ti--5Al has hysteresis values of 29 K and 15 K, respectively.  \footnote{They measured hysteresis using Diffraction Scanning Calorimetry (DSC), and by our method using Differential Thermal Analysis (DTA), the hysteresis values in these samples are found to be around 135 K and 116 K, respectively.}\\

The design criteria proposed in the literature should ideally be universally applicable to any new composition having dopants that are never explored before. In this study, we dope Zirconia with new dopants and design potential compositions that could reduce thermal hysteresis in SMCs based on the design criteria.
Dopants that reduce the tetragonality of the austenite phase towards a cubic crystal structure offer more degrees of freedom during transformation. 
This is because the phase transformation from the cubic to the monoclinic crystal structure has 24 variants of martensite, compared to only having 12 variants of martensite for the phase transformation from the tetragonal to monoclinic.
The availability of more variants in the cubic to monoclinic phase transformation makes this transformation more flexible, allowing them to accommodate high strain during transformation.
Hence, reducing the tetragonality of the austenite phase might lead to a low hysteresis SMC.
One possible dopant candidate is Erbia (Er$_2$O$_3$), which, when added to Zirconia, is known to reduce the tetragonality \cite{Sui2019} of the austenite phase. 
Therefore, we search for new samples in the compositional space of Erbia-doped Zirconia to check the universality of the design criteria.\\

To navigate the compositional space, we describe a data-driven methodology that identifies key physical features from electronic and crystal structure parameters. These structural parameters are obtained from the experimental X-ray diffraction data of multiple compositions. The physical features are selected as inputs for training a Gaussian process (GP) machine learning (ML) model based on their strong statistical correlation with ML outputs such as transformation temperatures and lattice parameters.
To search for a new SMC, a comprehensive library of synthetic compositions is generated by systematically sampling the space defined by the experimental minimum and maximum mole fraction boundaries of ceramic oxides. 
The GP models are then used to predict transformation temperatures and lattice parameters of the synthetic compositions. 
The predicted lattice parameters are used to calculate the maximum deviation from the exact satisfaction of the cofactor conditions and search for unique compositions that closely satisfy the cofactor conditions. \\

Our strategy identifies a composition ``31.75ZrO$_2$--37.75HfO$_2$--Y$_{0.5}$Ta$_{0.5}$O$_2$--1.5Er$_2$O$_3$'' whose predicted transformation temperature and lattice parameters are validated by our experimental work with high accuracy. 
Therefore, this strategy can be utilized to design multi-component ceramic systems for a desired transformation temperature and lattice parameters.
The composition 31.75ZrO$_2$--37.75HfO$_2$--14.5Y$_{0.5}$Ta$_{0.5}$O$_2$--1.5Er$_2$O$_3$ satisfies the design criteria proposed by Pang et al. \cite{Pang2022}.
However, it still has a significant thermal hysteresis of 137 \degree C during the differential thermal analysis (DTA) experiment.
This establishes that the criteria proposed by Pang et al. are not universal for every ceramic system, and there are other factors involved not yet known for achieving low hysteresis around 5 \degree C in SMCs.
%%%%%%%%%%%%%%%%%%%%%%%%%%%%%%%%%%%%%%%%%%%%%%%%%%%%%%%%%%%%%%%%%%%%%%%%%%%%%%%%%%%%%%%%%%%%%%%%%%%%%%%%
%%%%%%%%%%%%%%%%%%%%%%%---NEW SECTION---%%%%%%%%%%%%%%%%%%%%%%%%%%%%%%%%%%%%%%%%%%%%%%%%%%%%%%%%%%%%%%%%
%%%%%%%%%%%%%%%%%%%%%%%%%%%%%%%%%%%%%%%%%%%%%%%%%%%%%%%%%%%%%%%%%%%%%%%%%%%%%%%%%%%%%%%%%%%%%%%%%%%%%%%%

\section{Prediction of transformation temperature}
\label{section2:TT}
Transformation temperature is the temperature such that the change in total free energy between the austenite and the martensite phases of a system is zero, and the system is said to be in thermoelastic equilibrium.
The start and finish temperatures for forward transformation $M_\text{s}$ and $M_\text{f}$, and the temperatures for reverse transformation $A_\text{s}$ and $A_\text{f}$ play an important role in finding the range of
temperatures at which thermoelastic equilibrium occurs.
Tang and Wayman \cite{tong_characteristic_1974} found the range of thermoelastic equilibrium to be confined between $T_\text{o}$ and $T'_o$, where $T_\text{o} = \frac{1}{2}(M_\text{s}+A_\text{f})$ and $T'_o = \frac{1}{2}(M_\text{f}+A_\text{s})$.  
Thus, for any temperature $T$ between $T_\text{o}$ and $T'_o$ ($T_\text{o} > T > T'_o$), the thermoelastic equilibrium can be obtained. 
In this work, we have approximated the transformation temperature $T_\text{f}$ as the average of the four characteristic temperatures, i.e., $\frac{1}{4}(M_\text{s} + M_\text{f} + A_\text{s} + A_\text{f})$ \cite{zhang_energy_2009}. \\

To satisfy one of the design criteria of Pang et al., i.e., high transformation temperature ($M_\text{s} > 500$ \degree C) \cite{Pang2022},
selected dopants are added to ZrO$_2$ to raise its $M_s$.  
It is well known that alloying HfO$_2$ into ZrO$_2$ sharply increases the $A_\text{s}$ and $M_\text{s}$ temperatures \cite{Pang2022, Kim1990}. 
While HfO$_2$ slightly increases the tetragonality of the austenite phase in Yttria-stabilized Zirconia \cite{Kim1990}, it remains a preferred dopant because it elevates the $M_s$ more significantly than other oxides \cite{Pang2022}.
Additional dopants are also added that reduce the tetragonality of the austenite phase, such as $\text{Er}_2\text{O}_3$ and a mixture of Yttria and Tantalum pentoxide in a 1:1 ratio, i.e., Y$_{0.5}$Ta$_{0.5}$O$_2$ \cite{khor_lattice_1997, Kim1990}.
Lower mole fraction ($m$) values of $\text{Er}_2\text{O}_3$ \{$m_\text{$\text{Er}_2\text{O}_3$}|  0.01 \leq m_\text{$\text{Er}_2\text{O}_3$} \leq 0.03$ \} sharply decrease the tetragonality as compared to the same amount of Y$_{0.5}$Ta$_{0.5}$O$_2$ in ZrO$_2$.
While both $\text{Er}_2\text{O}_3$ and $\text{Y}_{0.5}\text{Ta}_{0.5}\text{O}_2$ reduce the $T_\text{f}$ of $\text{ZrO}_2$-based compositions, the $T_\text{f}$ can be maintained well above $500^\circ\text{C}$ by limiting the dopant concentrations to $m_{\text{Y}_{0.5}\text{Ta}_{0.5}\text{O}_2} \leq 0.18$ \cite{GURAK20183317} and $m_{\text{Er}_2\text{O}_3} \leq 0.03$ \cite{Duran_p_1977}.
Hence, we represent the ceramic system as (ZrO$_2$)$_{m_1}$--\,(HfO$_2$)$_{m_2}$--\,(Er$_2$O$_3$)$_{m_3}$--\,(Y$_{0.5}$Ta$_{0.5}$O$_2$)$_{m_4}$, where $m_4 = 1-({m_1}+{m_2}+{m_3})$.
We synthesize 44 compositions in which the mole fraction $m_j$ of each oxide is constrained as $0.1675 \leq m_1 \leq 0.71$, $0.1675 \leq m_2 \leq 0.5677$, $0.0 \leq m_3 \leq 0.055$ and $0.025 \leq m_4 \leq 0.175$.
During the synthesis process, the same processing conditions are applied to all 44 compositions to mitigate any effect of processing and microstructural changes.\\

These compositions are used to derive the set of key physical features based on electronic and crystal structure properties.
The features are used as input to a machine learning model that predicts $T_\text{f}$.
We compare key non-parametric ML methods for the model and choose the best-performing method for the predicting $T_\text{f}$.
The ML model capable of predicting $T_\text{f}$ plays a key role in searching for ceramic composition with targeted $T_\text{f}$.
Furthermore, the lattice parameters of the monoclinic and tetragonal crystal structures usually exhibit a significant correlation with $T_\text{f}$.
Therefore, the predicted $T_\text{f}$ by the ML model could also serve as an input parameter for the subsequent construction of another ML model for predicting lattice parameters. \\

In the next subsection, we discuss the set of key physical features and a way to derive the average value of the physical features for a given ceramic system.\\
%%%%%%%%%%%%%%%%%%%%%%%%%%%%%%%%%%%%%%%%%%%%%%%%%%%%%%%%%%%%%%%%%%%%%%%%%%%%%%%%%%%%%%%%%%%%%%%%%%%%%%%%
%%%%%%%%%%%%%%%%%%%%%%%---New Subsection---%%%%%%%%%%%%%%%%%%%%%%%%%%%%%%%%%%%%%%%%%%%%%%%%%%%%%%%%%%%%%
%%%%%%%%%%%%%%%%%%%%%%%%%%%%%%%%%%%%%%%%%%%%%%%%%%%%%%%%%%%%%%%%%%%%%%%%%%%%%%%%%%%%%%%%%%%%%%%%%%%%%%%%
\subsection{Search for correlated physical features}

We discuss here the importance of selecting the relevant features that form the input to the ML models.
Recently, there have been efforts in literature to understand a strong correlation between the $T_\text{f}$ and the physical features derived from the composition of the ceramic system.
For example, Zarinejad et al. \cite{Zarinejad2022} doped ZrO$_2$ with different oxides such as HfO$_2$, Y$_2$O$_3$, CeO$_2$, MgO, CaO and TiO$_2$ and depending on the chemical composition, found a clear correlation of the $T_\text{f}$ with the number of valence electrons ($ev$), valence electron ratio (\textit{VER}), and the atomic number ($Z$). 
They found that the $A_\text{s}$ and $M_\text{s}$ temperatures decrease with the increase in the \textit{VER} and the $ev$.
However, a clear positive correlation is observed between the $T_\text{f}$ and the average $Z$.
Similarly, Frenzel et al. \cite{frenzel_effect_2015} found a strong compositional dependence of $M_\text{s}$ in binary Ni-Ti, ternary Ni-Ti-Cr, and Ni-Ti-Cu alloys based on a strong stabilization of B2 austenite (resembling cubic cesium chloride type structure) through the formation of antisite defects. 
Thus, we also assume a likelihood of the formation of antisite defects influencing $T_\text{f}$ in ceramic oxides. Key factors causing antisite defects include stoichiometry, atomic size mismatch, and electronegativity.\\

Following the literature, we consider the following physical features that might strongly affect the $T_\text{f}$ in SMC: atomic number ($Z$), Clementi's atomic radius (\textit{rcle}), Pettifor chemical scale (\textit{cs}), 
Pauling electronegativity (\textit{en}), electron affinity (\textit{eff}), valence electrons (\textit{ev}), Shannon ionic radius (\textit{rio}), and  Slater atomic radius (\textit{rsla}). 
The average value of the physical features, $x$, for the composition (ZrO$_2$)$_{m_1}$--\,(HfO$_2$)$_{m_2}$--\,(Er$_2$O$_3$)$_{m_3}$--\,(Y$_{0.5}$Ta$_{0.5}$O$_2$)$_{m_4}$, are based upon a linear combination of weighted feature values of the cations in the composition. We use the mole fraction value $m_k$ of each oxide $k$ as the weight and represent $x$ in Eq. \ref{eq:featureX} as:

\begin{equation}
  \label{eq:featureX}
  \begin{aligned}
    x = \displaystyle \sum_{k=1}^{n_c} m_kp_k
  \end{aligned}
\end{equation}

Where the value of a physical feature for the cation of the oxide $k$ is represented by $p_k$ in Eq. \ref{eq:featureX}. The total no. of cations considered for creating the ceramic system is represented by $n_c$.
Valence electron ratio (\textit{VER}) is also considered to be one of the physical features in our study.
\textit{VER} is defined as the ratio of the average number of valence electrons to the average atomic number of the composition, defined as:

\begin{equation}
  \label{eq:featureVER}
  \begin{aligned}
    \textit{VER} = \frac{\displaystyle \sum_{k=1}^{n_c} m_k(\textit{ev})_k}{\displaystyle \sum_{k=1}^{n_c} m_k{Z}_k}
  \end{aligned}
\end{equation}

\begin{figure}[!ht]
  \centering
  \includegraphics[width=1\textwidth]{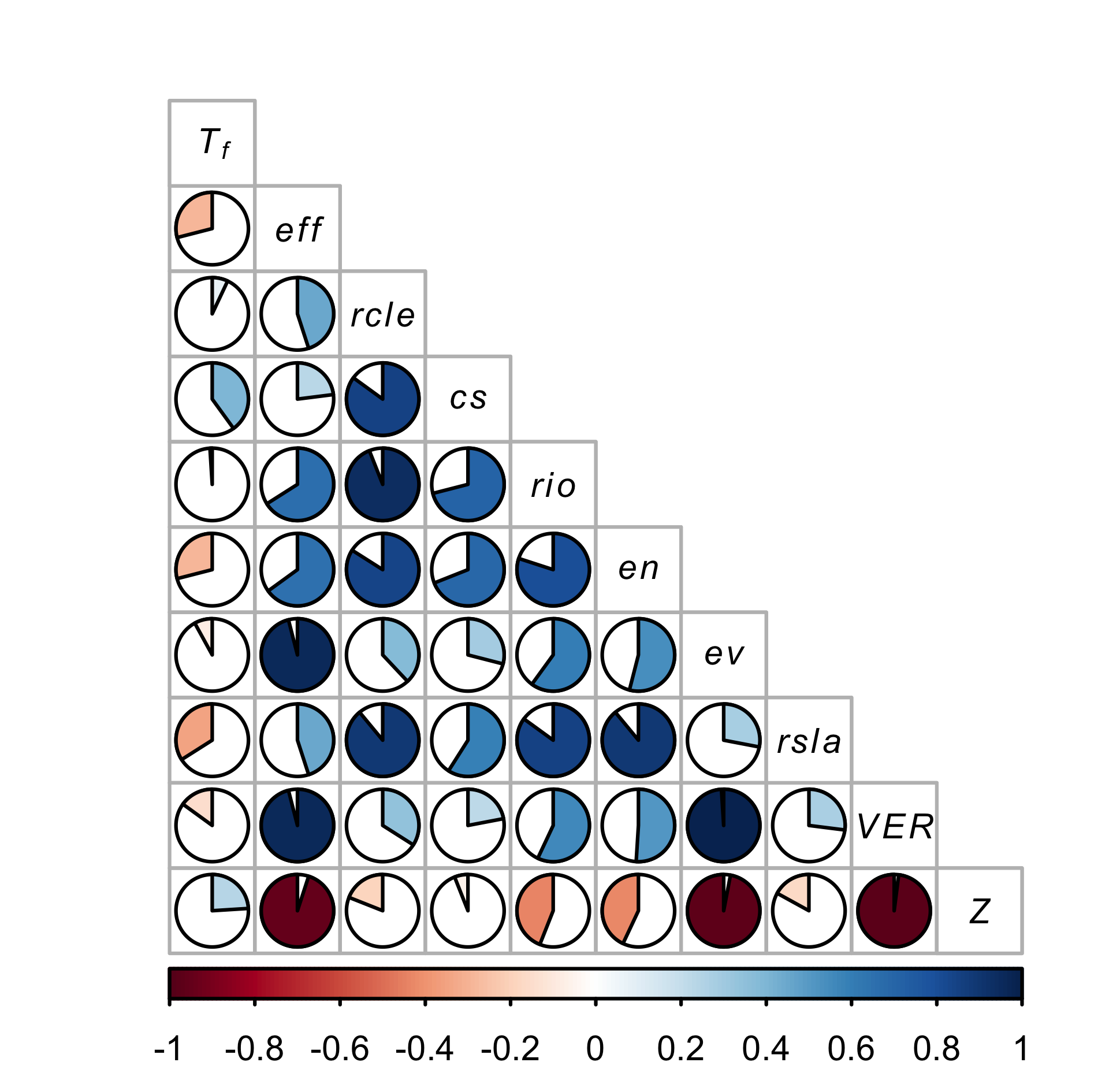}
  \caption[
      Pearson correlation map for features
  ]{
    Pearson correlation map for physical features: This graphical map shows Pearson cross-correlation coefficient among all the initial features and indicate the relative redundancies. Many features that have
    strong correlation among each other can be filtered out by just choosing one feature among all.
  }
  \label{fig1:Pearson map}
\end{figure}

Where $Z_k$ is the atomic number of cation that belongs to the oxide $k$. 
The set of physical features $\vec{x}$ derived from a ceramic system using Eq. \ref{eq:featureX} and Eq. \ref{eq:featureVER} are listed in Table \ref{table 1: List of features}.
Some of these features might strongly correlate with each other, leading to the use of redundant information as input to the ML model.
To identify the redundant features, we visualize the correlation among all features on Pearson correlation map \cite{Blyth1994} in Fig. \ref{fig1:Pearson map}, where a full blue circle indicates a +1 value of correlation. In comparison, a full red circle indicates a $-1$ value of correlation between variables.
We observe in Fig. \ref{fig1:Pearson map} that there exists a strong correlation among features \textit{Z}, \textit{ev}, \textit{VER}, \textit{eff}.
Since \textit{eff} has the highest correlation with $T_\text{f}$, we select the feature \textit{eff} and remove features \textit{Z}, \textit{ev}, \textit{VER} from the input space.
This results in only six features in the updated input space of the ML model, i.e., $\vec{x} =$ (\textit{rcle}, \textit{cs}, \textit{en}, \textit{rsla}, \textit{rio}, \textit{eff}). 
Building on this updated input space, the following section evaluates various ML regression models to identify the optimal architecture for predicting $T_\text{f}$. 

\begin{table}
 \caption{List of features considered in this work and their abbreviations}
  \centering
  \begin{tabular}{lll}
  \hline
    %\toprule
    %\multicolumn{2}{c}{Part}   \\
    %\cmidrule(r){1-2}
    Abbreviation     & Feature    \\
    \hline
    \midrule
    \textit{Z} &             Atomic number       \\
    \textit{rcle}    &       Clementi atomic radius (\AA) \textsuperscript{\cite{Clementi1967}}  \\
    \textit{cs}          &    Pettifor chemical scale\textsuperscript{\cite{Pettifor1985}} \\
    \textit{en}    &          Pauling electronegativity\textsuperscript{\cite{allred_electronegativity_1961}}\\
    \textit{ev}   &        Number of valence electrons\\
    \textit{rsla} &         Slater atomic radius \textsuperscript{\cite{Slater1964}}\\
    \textit{rio} &          Shannon ionic radius \textsuperscript{\cite{Shannon1976}}\\
    \textit{eff}  &         Electron affinity \textsuperscript{\cite{Hotop1985}}\\
    \textit{VER} &            Valence electron ratio \textsuperscript{\cite{Zarinejad2022}}\\
    \bottomrule
  \end{tabular}
  \label{table 1: List of features}
\end{table}

%%%%%%%%%%%%%%%%%%%%%%%%%%%%%%%%%%%%%%%%%%%%%%%%%%%%%%%%%%%%%%%%%%%%%%%%%%%%%%%%%%%%%%%%%%%%%%%%%%%%%%%%
%%%%%%%%%%%%%%%%%%%%%%%---New Subsection---%%%%%%%%%%%%%%%%%%%%%%%%%%%%%%%%%%%%%%%%%%%%%%%%%%%%%%%%%%%%%
%%%%%%%%%%%%%%%%%%%%%%%%%%%%%%%%%%%%%%%%%%%%%%%%%%%%%%%%%%%%%%%%%%%%%%%%%%%%%%%%%%%%%%%%%%%%%%%%%%%%%%%%
\subsection{Machine learning models for transformation temperature}\label{sssec:num2.2}
\begin{figure}[!ht]
  \centering
  \includegraphics[width=0.9\textwidth]{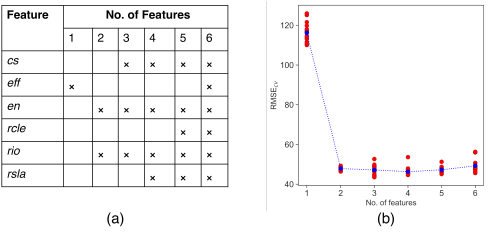}
  \caption[
      Effect of no. of features on RMSE
  ]{
    Effect of number of input parameters on RMSE of testing data obtained by Gaussian process regression (a) Six model has been tested with specific set of features 
    (b) RMSE of the testing data depends on the no. of features chosen in the model.
      }
  \label{fig2:features selection}
\end{figure}

The prepared experimental dataset for training the ML model could be represented as $$\mathcal{D} = \{(\vec{x}_r, T_{\text{f}, r})\}_{r=1}^{n}$$ for $n$ compositions, where $\vec{x}_r$ is the set of features for the $r^{th}$ composition.
The role of ML is to map the $d$--dimensional vector $\vec{x}_r$ into a singular scalar value $T_{\text{f},r}$ by learning a function $g(\vec{x}_r)$.
$$ g:\mathbb{R}^d \rightarrow \mathbb{R}$$
Our goal is to predict function values $g(\vec{x}^*)$ for the features $\vec{x}^*$ of the new synthetic compositions. 
Certain features within $\vec{x}$ may lack a consistent relationship with $T_\text{f}$, thereby contributing no predictive value. Consequently, it is essential to evaluate how feature selection impacts overall performance. To investigate this, we employed a Gaussian process (GP) regression model and created six GP models, each with a subset of features as input from $\vec{x}$.
The optimal subsets of features for each model were identified using best subset selection \cite{Tibshirani2016}, as illustrated in Fig. \ref{fig2:features selection}a.
In this approach, every possible combination of features for each subset is used as input to the model and the set of features that provides the greatest improvement in performance is selected.
The hyperparameters of these models are optimized using the RandomSearchCV \cite{Yoshua2012} algorithm.
The features of the best-performing model among the six GP models are used for the ML training and prediction.
\\

To measure the performance of GP models, we divide the full data $\mathcal{D} = \bigcup_{r=1}^{22} \mathcal{D}_r$ into 22 subsets, where the subsets $\mathcal{D}_r$ form a partition of the dataset $\mathcal{D}$, such that $\mathcal{D}_i \cap \mathcal{D}_j = \emptyset$ for all $i \neq j$.
Among 22 subsets, a unique subset is held out for testing while the remaining 21 subsets are used for training.
This process is known as K-fold cross-validation and for our model, it is 22-fold cross-validation.
The GP model predicts $T_\text{f}$ for each testing subset, and these predictions are combined to form a single array of predicted $T_\text{f}$ for all 44 compositions.
The performance of the cross-validation is evaluated based on the error matric root mean square error ($\text{RMSE}$) between the predicted $T_\text{f}$ and actual $T_\text{f}$ of all 44 compositions.
The cross-validation is repeated 20 times to average out the variability in the performance of $\text{RMSE}$. 
The repeated cross-fold validation is plotted for all six GP models, and shown in Fig. \ref{fig2:features selection}b.
Here, we find that the GP model with four features (\textit{cs}, \textit{rio}, \textit{en}, \textit{rsla}) has the minimum value of the mean $\text{RMSE}$.
This highlights that the addition of features \textit{rcle} and \textit{eff} in the fifth and sixth GP models do not improve the mean value of $\text{RMSE}$.
Hence, the four-feature model was selected for training and the subsequent prediction of $T_\text{f}$.\\

Recent literature features a diverse array of non-parametric machine learning methods for modeling structure-property relationships in materials.
For example, Liu et al. \cite{LIU2021100898} utilized a variety of physical features as model inputs, including elemental properties, reactivity, thermal characteristics, and electronic structure configurations. They used Support Vector Regression (SVR), Random Forest (RF) and Gaussian Process (GP) model to predict
$T_\text{f}$ of NiTiHf shape memory materials.
They concluded that the GP model not only provides superior accuracy in response prediction but also estimates variance in response (i.e., uncertainty in prediction).
Knowledge of uncertainty in the response prediction gives a confidence interval of prediction, which is useful in identifying a synthetic composition with confidence for future XRD experiments.
Similarly, Kankanamge et al. \cite{Kankanamge2022} used alloy composition of NiTiHf shape memory alloys and predicted its martensite start temperature $M_\text{s}$ using linear, polynomial, SVR, and K-nearest neighbor (KNN) and found that the KNN model shows best performance in prediction of $M_\text{s}$.\\

Based on the literature survey, we selected several non-parametric ML algorithms, namely RF, GP, and KNN to compare their performance in predicting $T_\text{f}$ for SMC.
Each algorithm takes the four features (\textit{cs}, \textit{rio}, \textit{en}, \textit{rsla}) as input and predicts $T_\text{f}$ as output.
We implement the 22-fold cross-validation for each algorithm.     
The training subset from the last cross-validation fold is used to check the performance in predicting training data. In this case, RF predicts with superior performance than GP and KNN.
However, when predicted $T_\text{f}$ for each test subset of the cross-validation are combined to measure a collective test performance, GP outperforms KNN and RF.
The performance matrices--coefficient of determination ($\text{R}^2$) and RMSE--for the training and testing data are shown in Table \ref{table2:performance_matric_for_Tf}.

% Define a custom strut command for row height
\newcommand{\customstrut}{\rule{0pt}{1cm}}
\begin{table}[h!]
\centering
\caption{R\textsuperscript{2} and RMSE for GP, KNN and RF}
\begin{tabular}{|c|c|c|c|c|}
\hline
\customstrut \textbf{Model} & \parbox{1cm}{\textbf{R\textsuperscript{2}} \\ Train\\} & \parbox{1cm}{\textbf{R\textsuperscript{2}} \\ Test\\} & \parbox{2cm}{\textbf{RMSE} \\ Train\\} & \parbox{2cm}{\textbf{RMSE} \\ Test\\} \\
\hline
Gaussian Process (GP) & 0.92 & 0.85 & 34.17 & 49.30 \\
\hline
K-nearest neighbour (KNN) & 0.87 & 0.77 & 44.72 & 59.60\\
\hline
Random Forest (RF) & 0.96 & 0.74 & 22.69 & 62.82 \\
\hline
\end{tabular}
\label{table2:performance_matric_for_Tf}
\end{table}

The predicted $T_\text{f}$ for all the test subsets are compared with the actual $T_\text{f}$ by GP, KNN, and RF models in Fig. 3(a), (b), and (c) respectively.
Here, as measured by the R$^2$ metric, the GP, KNN, and RF models explain 85\%, 77\%, and 74\% of the variance in the predicted $T_\text{f}$, respectively.
Additionally, GP provides confidence in its prediction in terms of uncertainty, known as Epistemic uncertainty. The uncertainty bands are plotted in Fig. 3(a) as $T_\text{f} \pm 1.96\sigma$, where $\sigma$ is the predictive standard deviation. This indicates a 95\% probability that the true $T_\text{f}$ value for any given input $\vec{x}_r$ falls within the interval of $T_\text{f} \pm 1.96\sigma$.

%---------------------------------------------------------------------------

\begin{figure}[!ht]
  \centering
  \includegraphics[width=1\textwidth]{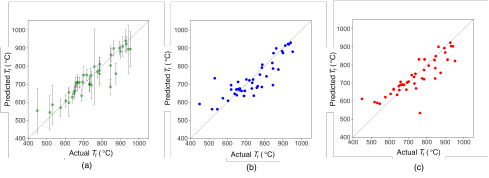}
  \caption[
      Performance of ML models on testing data of $T_f$
  ]{
    Performance of ML models on all test subsets: Comparison of Actual vs. Predicted $T_\text{f}$ by (a) GP (b) KNN and (c) RF. Figure 3(a) also contains the uncertainty in prediction as error bars.
      }
  \label{fig:correspondence:1ab}
\end{figure}

%%%%%%%%%%%%%%%%%%%%%%%%%%%%%%%%%%%%%%%%%%%%%%%%%%%%%%%%%%%%%%%%%%%%%%%%%%%%%%%%%%%%%%%%%%%%%%%%%%%%%%%%%%%%
%NEW SECTION%
%%%%%%%%%%%%%%%%%%%%%%%%%%%%%%%%%%%%%%%%%%%%%%%%%%%%%%%%%%%%%%%%%%%%%%%%%%%%%%%%%%%%%%%%%%%%%%%%%%%%%%%%%%%%
\section{Prediction of lattice parameters} \label{sec:nam3}

A robust ML model that predicts lattice parameters of crystal structures would enable calculation of cofactor conditions \cite{Chen2013} and identify compositions that closely satisfy cofactor conditions.
The lattice parameters of both monoclinic and tetragonal crystals should be predicted with very high accuracy for close satisfaction of the cofactor conditions.
We again chose the GP model for prediction of the lattice parameters due to its superior performance and ability to provide predictions as a normal distribution.
The GP models for the monoclinic lattice parameters is termed GP$_\text{m}$ and for the tetragonal lattice parameters is termed as GP$_\text{t}$.
The key physical features as input for the GP$_\text{m}$ and GP$_\text{t}$ model are identified in the next section.

\subsection{Predictions of monoclinic crystal's lattice parameters}
In the monoclinic crystal structure, the conventional unit cell of the lattice has the lattice parameters $a_\text{m}$, $b_\text{m}$, $c_\text{m}$ and the angle between them are $90 \degree$, $\beta$, $90 \degree$.
The GP$_\text{m}$ models predict $a_\text{m}$, $b_\text{m}$, $c_\text{m}$, and $\beta$.
The lattice parameters are strongly dependent on changes in temperature ($T$) \cite{GURAK20183317} of the crystal, thus $T$ is one of the key input feature in both the GP$_\text{m}$ and GP$_\text{t}$.
The Pearson correlation coefficient between the lattice parameters and $T$ is calculated  based on the experimental X-ray diffraction data and shown in Table \ref{table3:pearsoncorr_lp}. 
Here, the lattice parameters $a_\text{m}$, $c_\text{m}$ and $\beta$ have strong correlation with $T$, however, $b_\text{m}$ has weak correlation with $T$.
Consequently, only $T$ is not a sufficient input to the GP$_\text{m}$ for high accuracy in prediction of $b_\text{m}$.
Thus, we consider including addition physical features from Fig. \ref{fig2:features selection}a to the input space of GP$_\text{m}$.   
The relevant features necessary for high prediction accuracy of the GP$_\text{m}$ are identified based on features selection study to predict $b_\text{m}$.\\

\begin{table}[h!]
\centering
\caption{Pearson correlation coefficient between $T$ and lattice parameters}
\begin{tabular}{|c|c|c|c|c|c|c|c|}
\hline
\strut    & \parbox{1cm}{$a_m$} & \parbox{1cm}{$b_m$} & \parbox{1cm}{$c_m$} & \parbox{1cm}{$\beta$} & \parbox{1cm}{$a_t$} & \parbox{1cm}{$c_t$} & $T$\\
\hline
$T$ & 0.92 & - 0.07 & 0.86 & -0.82 &  0.92 & 0.61 & 1 \\
\hline
\end{tabular}
\label{table3:pearsoncorr_lp}
\end{table}

\begin{figure}[!ht]
  \centering
  \includegraphics[width=1\textwidth]{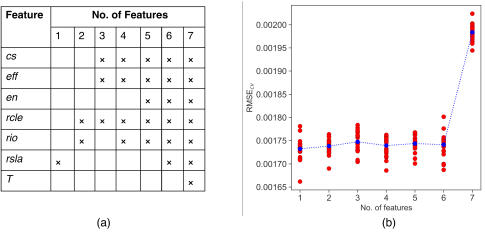}
  \caption[
      Performance of ML models on $b_m$ testing data
  ]{Effect of no. of input parameters on RMSE of testing data obtained by GP regression (a) Seven GP regression models with best subset of features for prediction of lattice parameters $b_m$ (b) RMSE of the testing data depends on the no. of features.
      }
  \label{fig4:No_of_features_for_bm}
\end{figure}

In the feature selection study, our goal is to identify key physical features among seven features ($T$, \textit{rcle}, \textit{cs}, \textit{en}, \textit{rsla}, \textit{rio}, \textit{eff}) to represent the input space of the GP models to predict $b_\text{m}$.
For this study, seven GP models are created, with the first one having only one feature and each subsequent model having an additional single increment in the number of features, as shown in Fig. \ref{fig4:No_of_features_for_bm}a.
The best input features for a model are obtained by the best subset selection \cite{Tibshirani2016}. 
Each model was evaluated using 20-fold cross-validation with 20 repetitions, as shown in Fig. \ref{fig4:No_of_features_for_bm}b.
We observe that the mean value of $\text{RMSE}_{\text{cv}}$ does not significantly change by including up to 4 features ($S_l=$(\textit{eff}, \textit{rcle}, \textit{cs}, \textit{rio})).
The GP model with 4 features also gives the least variance in the RMSE$_\text{cv}$ value among all 7 GP models, marking $S_l$ as the best subset.\\

\begin{figure}[!ht]
  \centering
  \includegraphics[width=1\textwidth]{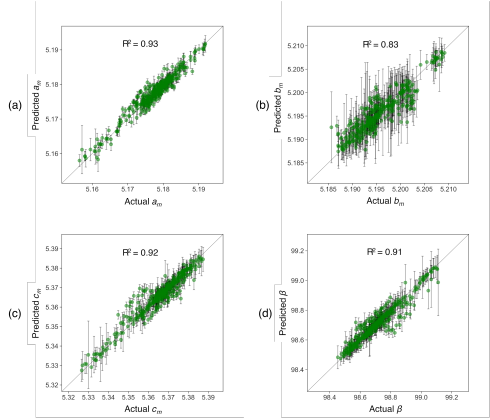}
  \caption[
      Predictions of Monoclinic lattice parameters
  ]{Predicted vs. actual values of monoclinic crystal's lattice parameters: (a) $a_\text{m}$ (b) $b_\text{m}$ (c) $c_\text{m}$ (d) $\beta$. 
      }
  \label{fig6:monoclinic lattice params}
\end{figure}

We also include the $T$ in the current subset because the lattice parameters $a_\text{m}$, $c_\text{m}$ and $\beta$ have a strong correlation with the $T$.
Thus, $S_l$ is updated as $S_{lm}=($T$, \textit{eff}, \textit{rcle}, \textit{cs}, \textit{rio})$ and it becomes input to predict all monoclinic lattice parameters.
It makes physical sense for GP$_\text{m}$ to have common input $S_{lm}$ and predict all lattice parameters of the monoclinic crystal. The lattice parameters $a_\text{m}$, $b_\text{m}$, $c_\text{m}$ and $\beta$ are taken as output individually during training the GP$_{\text{m}}$. The predictions of lattice parameters $a_\text{m}$, $b_\text{m}$, $c_\text{m}$ and $\beta$ are shown in Fig. \ref{fig6:monoclinic lattice params}a, b, c, and d respectively.\\

\subsection{Predictions of tetragonal crystal's lattice parameters}
\begin{figure}[!ht]
  \centering
  \includegraphics[width=1\textwidth]{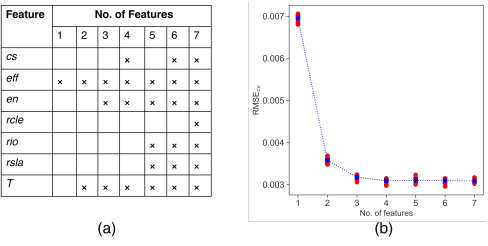}
  \caption[
     Feature selection for tetragonal lattice parameters
  ]{Study of feature selection for the tetragonal lattice parameters: (a) Seven GP models, each with best features for minimizing RMSE (b) RMSE of the testing data depends on the no. of selected features.}
  \label{fig6:featuresstudy_tetragonal}
\end{figure}

In the tetragonal crystal structure, the conventional unit cell of the lattice has the lattice parameters a$_\text{t}$ = b$_\text{t}$ and c$_\text{t}$. The angle between them are all $90 \degree$.
The GP$_\text{t}$ models predict a$_\text{t}$ and c$_\text{t}$.
To identify key features among seven features $T$, \textit{rcle}, \textit{cs}, \textit{en}, \textit{rsla}, \textit{rio}, \textit{eff} as inputs in the GP$_\text{t}$ model, the feature selection study is conducted.
As shown in Table \ref{table3:pearsoncorr_lp}, $c_\text{t}$ has a weaker relative correlation of 0.61 with $T$ compared to $a_\text{t}$, which shows a correlation of 0.92.
Hence, the lattice parameter c$_\text{t}$ is predicted using GP models.
We follow a feature selection process similar to the one used for the monoclinic lattice parameters; the resulting performance of seven GP models, each incorporating an increasing number of features, is illustrated in Fig. \ref{fig6:featuresstudy_tetragonal}a. 
\\

\begin{figure}[!ht]
  \centering
  \includegraphics[width=1\textwidth]{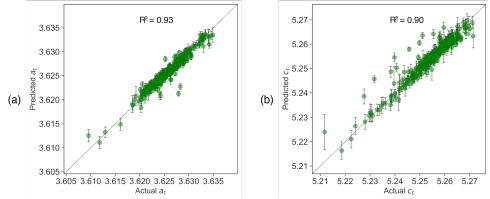}
  \caption[
     Feature selection for tetragonal lattice parameters
  ]{Predicted vs. actual values of tetragonal crystal's lattice parameter (a) $a_\text{t}$ (b) $c_\text{t}$}
  \label{fig7:tetragonal_predictions}
 
\end{figure}

The mean value of the RMSE$_\text{cv}$ of the 20 times repeated 20-fold cross validation does not significantly change after including 4 features ($T$, \textit{eff}, \textit{cs}, \textit{en}) in the model, as shown in Fig. \ref{fig6:featuresstudy_tetragonal}b.
The features that provide high efficiency in the prediction of the lattice parameter c$_\text{t}$ are likely to be informative for predicting the lattice parameter a$_\text{t}$.
Thus, it makes physical sense to use these four features as input for prediction of both the $a_\text{t}$ and $c_\text{t}$ individually by the GP$_\text{t}$.
The predictions from all the testing sets of the cross-fold validation are combined and compared with actual XRD measurements in Fig. \ref{fig7:tetragonal_predictions}.\\

%%%%%%%%%%%%%%%%%%%%%%%%%%%%%%%%%%%%%%%%%%%%%%%%%%%%%%%%%%%%%%%%%%%%%%%%%%%%%%%%%%%%%%%%%%%%%%%%%%%%%%%%%%%%
%NEW SECTION%
%%%%%%%%%%%%%%%%%%%%%%%%%%%%%%%%%%%%%%%%%%%%%%%%%%%%%%%%%%%%%%%%%%%%%%%%%%%%%%%%%%%%%%%%%%%%%%%%%%%%%%%%%%%%
\section{Theory of reversibility of phase transformation}\label{sec:cofactorcondition}

In this section, we review conditions of compatibility between the phases with regard to reversibility and hysteresis of phase transformation.
These conditions are purely geometric and depends only on the crystal structure and lattice parameters of the two phases.
These conditions influence the stress in transition layers and the heights of energy barriers that relate to hysteresis.
We will describe two conditions of supercompatibility, $\lambda_2=1$ and the \textit{cofactor conditions}.\\

The \textit{transformation stretch matrix} and the two groups that represent the point group symmetries of the two phases, are the terms used to express the two conditions of supercompatibility. 
In order to define the transformation stretch matrix, we first observe that in the most reversible martensitic phase transformations, the point group symmetries of the two phases have a group–subgroup relation.
The point group of the low temperature phase is a subgroup of the point group of the high temperature phase.
Hence, there exists a primitive lattice describing the periodicity of the martensite crystal, as delineated by vectors $\vec{b}_\text{1}$, $\vec{b}_\text{2}$, $\vec{b}_\text{3}$, and a sublattice of the austenite with periodicity $\vec{a}_\text{1}$, $\vec{a}_\text{2}$, $\vec{a}_\text{3}$ exhibiting about the same unit-cell volumes.
During phase transformation, the martensite primitive lattice is deformed from the austenite sublattice through a linear transformation $\mat{F}$ ($\det \mat{F} \neq 0$) such that $\vec{b}_i = \mat{F}\vec{a}_j$.
If $\mat{F}$ is assumed to have positive determinant by changing sign of one of the vectors, then, $\mat{F}$ has a polar decomposition $\mat{F} = \mat{RU}$, where $\mat{R}$ is a $3 \times 3$ rotation matrix, and $\mat{U}$ is the \textit{transformation stretch matrix}, which is positive definite and symmetric.\\

The primitive lattice of the martensite may have about the same unit cell volume as several sublattices of the austenite.
Based on study by Bain \cite{Bain1924} and Lomer \cite{lomer_beta_nodate}, it is observed that given the sublattice of the austenite $\vec{a}_\text{1}$, $\vec{a}_\text{2}$, $\vec{a}_\text{3}$, the material often chooses the primitive lattice of martensite which gives the smallest strain $||\mat{U}- \mat{I}||$, measured in a suitable norm.
Chen et al. \cite{Chen2013} and Koumatos et al. \cite{koumatos_optimality_2016} describe suitable algorithms for calculating $\mat{U}$ based on this principle.
This procedure shares a close relationship with the widely used \textit{Cauchy-Born rule}.
This rule is used to link atomic level deformations to macroscopic deformation.
A continuous deformation $\bm{y(x)}$ defined on a domain $\Omega$ has a gradient that takes the value ${\nabla \vec{y}}$. 
The interpretation of this rule is, ${\mat{F} = \nabla \vec{y}}$ represents the macroscopic deformation gradient.\\ 

The $\mat{U}$ has three real eigenvalues $0 < \lambda_1 \leq \lambda_2 \leq \lambda_3$, since it is positive definite and symmetric.
Among these eigenvalues, $\lambda_2$ is of particular importance because of the following theorem \cite{ball_fine_1987} (Prop. 4):\textit{a necessary and sufficient condition that a continuous deformation ${\vec{y}(\vec{x})}$ defined on a domain $\Omega$ has a gradient that takes value ${\nabla \vec{y} = \mat{F} = \mat{RU}}$ (martensite) on a region $\mathcal{R}$ and $\nabla \vec{y} = \mat{I}$ (austenite) on the complementary region $\Omega \backslash \mathcal{R}$ for some rotation matrix $\mat{R}$ is that $\lambda_2 = 1$.}
To relate this statement to Prop. 4 of \cite{ball_fine_1987}, note that ${\vec{y}(\vec{x})}$ of this form is continuous if and only if $\mat{F} = \mat{RU} = \mat{I} + \vec{b} \otimes \vec{m}$ for some vectors $\vec{b}$ and $\vec{m}$.
The vector $\vec{m}$ is taken as normal to the interface between the austenite and the martensite and the vector $\vec{b}$ is the shape strain vector. 
We take $\mat{F}^{\text{T}}\mat{F}$ to eliminate the rotation matrix $\mat{R}$ and note that the eigenvalues of $\mat{C} = \mat{F}^{\text{T}}\mat{F}$ are the squares of the eigenvalues of the positive definite, symmetric matrix $\mat{U}$ so, in particular, $\mat{U}$ has middle eigenvalue equal to 1 if and only if $\mat{C} = \mat{F}^{\text{T}}\mat{F}$ has the middle eigenvalue equal to 1.\\

The compatibility condition $\lambda_2 = 1$ strongly affects hysteresis \cite{zarnetta_identification_2010, bucsek_composition_2016, cui_combinatorial_2006, zhang_energy_2009}, and improves reversibility under thermal cycling \cite{zarnetta_identification_2010}.
A detailed theory of influence of geometric condition on hysteresis is given in \cite{zhang_energy_2009}.
The summary of the theory is as follows.
This is observed in both theory \cite{zhang_morphology_2007} and experiment \cite{fisher_nucleation_1948} that a well-developed nuclei of martensite exist above the austenite finish temperature $A_\text{f}$.
The theory \cite{zhang_energy_2009}, based on a concept of metastability, hypothesizes that lowering the temperature below that at which two bulk phases have the same free energy leads to growth of a twinned platelet.
During small undercooling, a spontaneous thickening of the platelet results in an increase in energy. 
This is attributed to the subtle interplay between bulk and interfacial energy at the twinned austenite/martensite interfaces that bind the platelet.
The bulk energy dominates at sufficiently large sizes of the platelet, and growth of the platelet leads to a decrease in the energy with size.
The large-scale transformation is caused at a certain under-cooling when a sufficient number of nuclei have a size beyond the barrier. \\

\begin{figure}[!ht]
  \centering
  \includegraphics[width=1\textwidth]{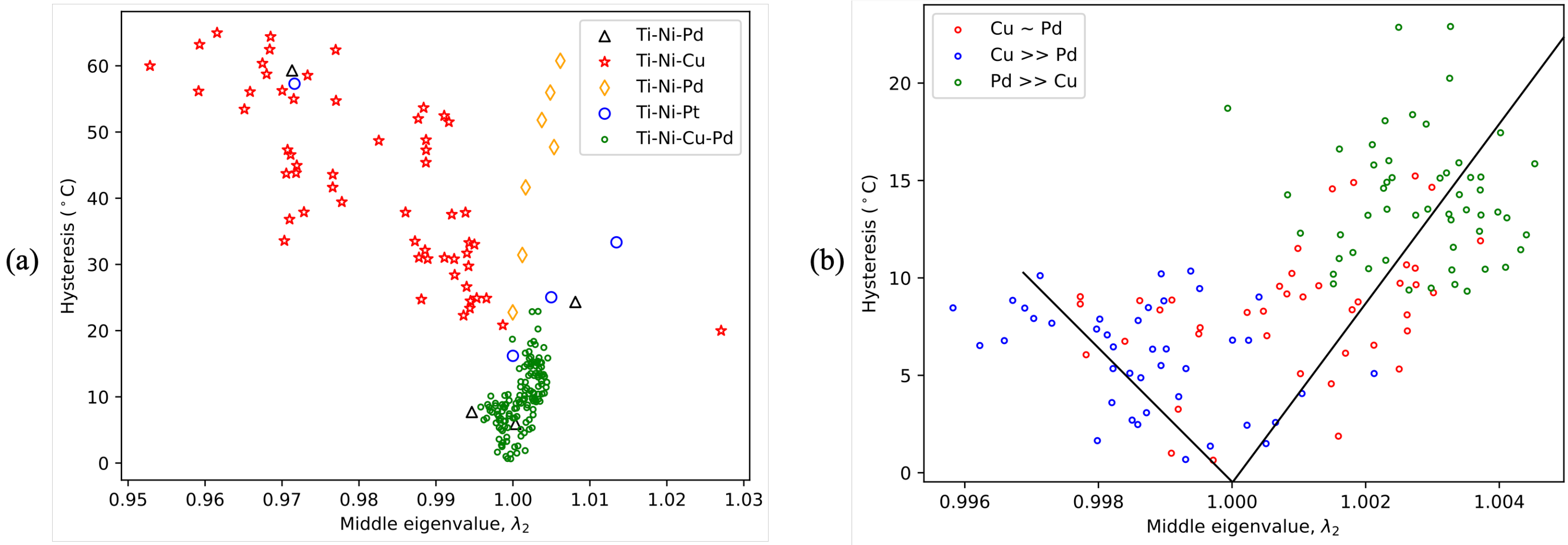}
  \caption[
     plot of lambda2 vs hysteresis
  ]{Measurements for alloys in Ti-Ni-X system (a) Hysteresis vs $\lambda_2$ (b) Closeup of the data in (a) centered near $\lambda_2 = 1$, taken from \cite{zarnetta_identification_2010}.}
  \label{fig_lambda2_vs_hys}
\end{figure}

The delicate interplay between bulk and interfacial energy at the austenite/martensite interface is the main reason why $\lambda_2$ so strongly affects hysteresis.
At $\lambda_2 = 1$, the existence of a perfectly unstressed interface between phases implies that the elastic energy in the stressed transition layer is eliminated.
As $\lambda_2$ departs from 1, and particularly in the case when $\lambda_2 > 1$, the bulk energy in the transition layer grows extremely rapidly.
When this theory is applied to cubic to orthorhombic transformations in TiNiX alloys, a graph of hysteresis vs. $\lambda_2$ is observed similar to that shown in Fig. \ref{fig_lambda2_vs_hys}. 
A very sharp drop in hysteresis is observed near $\lambda_2 = 1$ as shown in Fig. \ref{fig_lambda2_vs_hys}(b).

\subsection{Cofactor conditions}\label{sec:cofactor}
The cofactor conditions \cite{james_way_2005, Chen2013} consist of three conditions (i) the compatibility condition $\lambda_2 = 1$ discussed above (ii) a second condition that depends on the twin system chosen, and (iii) an inequality that is satisfied for Type-I or Type-II twin system. 
In this section, we discuss meaning of satisfying (ii) only.\\

The cofactor conditions are derived from the crystallographic theory of martensite \cite{bowles_crystallography_1954, WECHSLER1953}. 
The ubiquitous austenite/twinned martensite interface (``habit plane'') in martensitic phase transformation is governed by this theory.
By refining the twins, it provides necessary and sufficient conditions for the bulk energy in the elastic transition layer between phases to become vanishingly small.
This theory has as unknowns the volume fraction $f$ of twins in the laminate of the martensite phase defined in the domain $\Omega$, a rigid body rotation $\mat{R}_\text{a}$ of the austenite, a unit normal $\bm{m}$ to the habit plane and a shear vector $\bm{b}$ \cite{ball_fine_1987}.
We summarize the theory in brief and say that the domain $\Omega$ has a martensite laminate with the twin structure.
Let's say that the twin structure has two variants, where deformation gradient $\mat{R}_1\mat{U}_i$ represents one of the variants having volume fraction $f$, and the other variant is represented by the deformation gradient $\mat{R}_2\mat{U}_j$ having volume fraction $(1-f)$.
For the deformation $\vec{y}(\vec{x})$ to be continuous in the twin laminate, it is necessary that the deformation gradients $\mat{R}_1\mat{U}_i$ and $\mat{R}_2\mat{U}_j$ satisfy $\mat{R}_1\mat{U}_i - \mat{R}_2\mat{U}_j = \vec{a} \otimes \vec{n}$.
The interface between the two region is a plane with reference normal $\vec{n}$, and $\vec{a}$ is the twinning shear vector.
This condition is known as kinematic compatibility condition.\\

The average deformation gradient for the martensite laminate is $\mat{F}_\text{a} = f\mat{R}_1\mat{U}_i + (1-f)\mat{R}_2\mat{U}_j$.
The polar decomposition of the $\mat{F}_\text{a} = \mat{R}_\text{a}\mat{U}_\text{a}$, where $\mat{R}_{\text{a}}$ is the rotation of the austenite and $\mat{U}_\text{a}$ is the average transformation stretch matrix.
The deformation $\vec{y}(\vec{x})$ defined on the domain $\Omega$ is continuous if and only if $\mat{R}_\text{a}\mat{U}_\text{a} = \mat{I} + \bm{b} \otimes \bm{m}$.
Note that the eigenvalues of $\mat{C}$ are the squares of the eigenvalues of the positive-definite, symmetric matrix $\mat{U}_\text{a}$.
Now, $\mat{U}_\text{a}$ has the middle eigenvalue $\lambda_2=1$ if and only if $\mat{C} = \mat{F}_\text{a}^{\text{T}}\mat{F}_\text{a} = \mat{U}_\text{a}^\text{T}\mat{U}_\text{a}$ has the middle eigenvalue equal to 1.
This implies that one of the roots of the characteristic polynomial det$(\mat{C}-\mat{I})$ must be zero, which is $(\lambda_2^2-1)=0$.\\

\begin{figure}[!ht]
  \centering
  \includegraphics[width=0.9\textwidth]{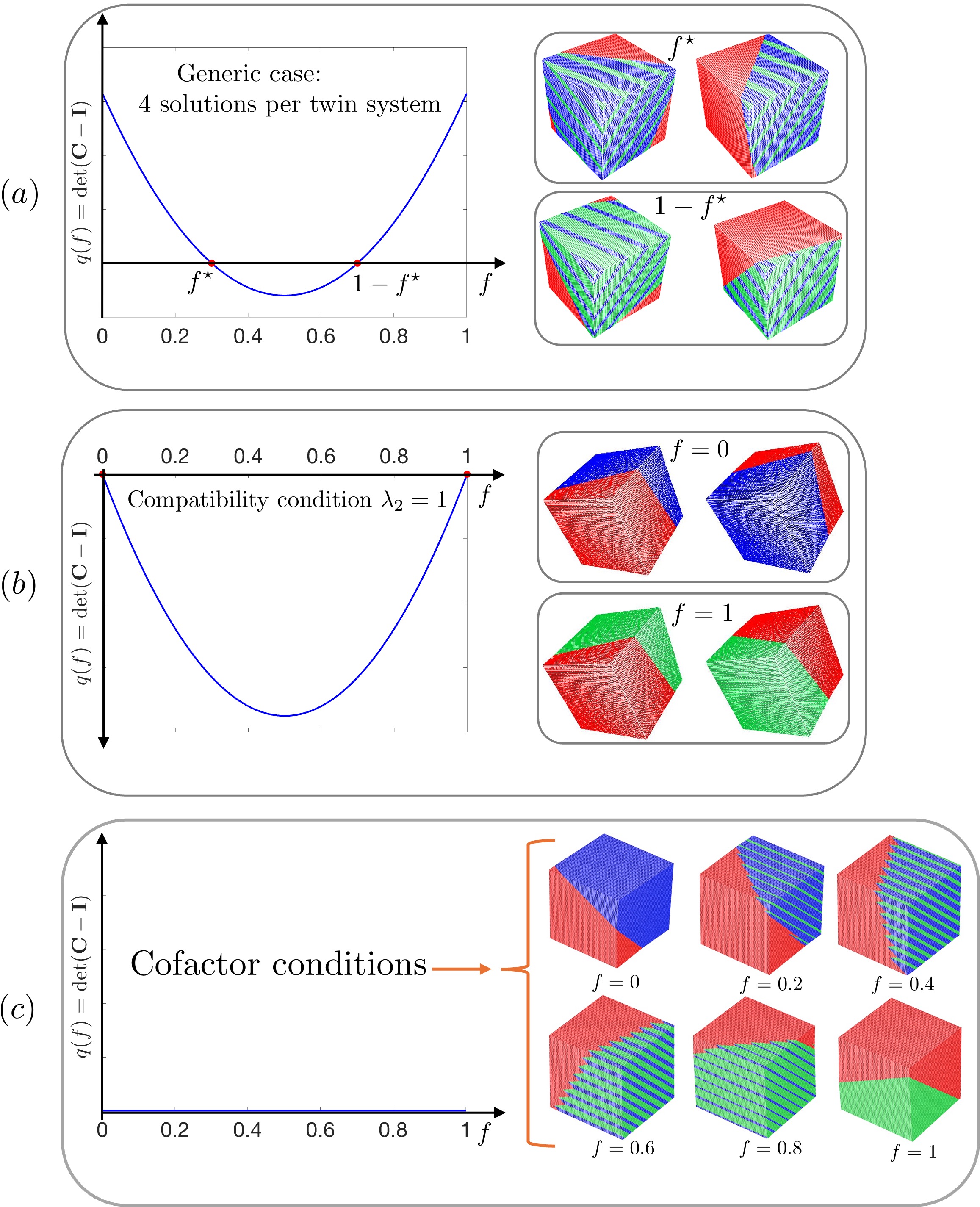}
  \caption[
     plot of lambda2 vs hysteresis
  ]{In the context of the crystallographic theory of martensite, the meaning of cofactor condition is: (a) No roots of the $\text{det}(\mat{C} - \mat{I})$ leading to no solution for the given twin system (b) the generic case of 4 solution per twin system satisfied by many reversible martensites (c) the case $\lambda_2 = 1$ (d) the cofactor conditions are exactly satisfied (The accompanying illustration depicts an example of type-I twins). Color code: blue and green are variants of martensite, and red is austenite.}
  \label{fig_meaning_cofactor}
\end{figure}

This implies that the theory reduces to a single scalar equation $\text{det}(\mat{C} - \mat{I}) = 0 $ for the volume fraction $f$ (See Theorem 2 of \cite{Chen2013} for the definition of $\mat{C}$).
It turns out \cite{ball_fine_1987} that in all cases, $\det(\mat{C}-\mat{I}) = 0$ is a quadratic equation for $0 \leq f \leq 1$. 
There are various solutions of this quadratic equation, as seen in Fig. \ref{fig_meaning_cofactor}. 
This may have no real roots, as seen in many martensitic steels that exhibit non-reversible martensite.
Alternatively, it may have exactly two roots, $f^*$ and $(1-f^*)$, as shown in Fig. \ref{fig_meaning_cofactor}a; many NiTi- or Cu-based shape memory alloys satisfy this classic case.
Also, a mild inequality must be satisfied for these roots to give a solution.
If it holds, then among these two roots, there are two interfaces corresponding to $f^*$ and two interfaces corresponding to $(1-f^*)$ i.e., four solution per twin system as shown in Fig. \ref{fig_meaning_cofactor}a(right). 
If these two roots $f^*$ and $(1-f^*)$ occurs at 0 and 1, as shown in Fig. \ref{fig_meaning_cofactor}b, and again if a certain inequality holds, this is the situation we discussed in the Sec. \ref{sec:cofactorcondition} above for the compatibility condition $\lambda_2 = 1$.
The four possible interfaces are shown to the right of Fig \ref{fig_meaning_cofactor}b.
Finally, if the quadratic function is identically zero for all values of $f$ as shown in Fig. \ref{fig_meaning_cofactor}c, then these conditions are called \textit{cofactor conditions}.
Assuming again a certain inequality is usually satisfied, this means that there exist low energy interfaces for any volume fraction $ 0 \leq f \leq 1 $ of the twins.\\

There are hosts of implications of satisfying the cofactor conditions \cite{Chen2013}. 
Satisfaction of cofactor conditions for one twin system implies its satisfaction for other crystallographic twin system.
Also, the solution of the crystallographic theory for Types I and II twin system exist with no elastic transition layer, platelet nucleation mechanism with zero elastic energy, and complex ``riverine'' zero energy microstructure \cite{song_enhanced_2013}.
There are strong correlations \cite{gu_phase_2018} that relates the satisfaction of cofactor condition to reversibility.
These correlations are possible on satisfaction of cofactor condition due to many strain and many interfaces possible in low (or zero)-energy microstructure involving both austenite and martensite.\\

The quadratic function $q(f) = \text{det}(\mat{C} - \mat{I})$ vanishes identically if and only if 
$q(0)=0$ and $q^{\prime}(0)=0$.
In terms of the twin system $\bm{a}$, $\bm{n}$ (Types I and II) and transformation stretch matrix $\mat{U}$, the cofactor conditions \cite{Chen2013} are:
\begin{align*}
& q(0)  = 0 \longleftrightarrow \lambda_2 = 1\\
& q^{\prime}(0) = 0 \longleftrightarrow \vec{a}.\mat{U}_j\text{cof}(\mat{U}_j^2-\mat{I})\bm{n}\\
& \quad \quad = 0 ,\text{CCI (for Type I twin) or CCII (for Type II twin)} \\
& \mathrm{tr}\,\mat{U}_j^2 - \det \mat{U}_j^2 - \frac{\bm{a}^2\bm{n}^2}{4} - 2 \geq 0
\end{align*}
The latter is an inequality referred to above. It is satisfied for all Types I and II twin system.\\

\subsection{Multiple transformation correspondences}
In tetragonal to monoclinic phase transformation, Kelly et al. \cite{Kelly2002} pointed out the presence of three types of correspondences, which describes how the basis vectors of the tetragonal lattice are mapped to the monoclinic lattice. 
These correspondences are also noticed by Hayakawa et al. \cite{Hayakawa1989I, Hayakawa1989II} in $\text{ZrO}_2$-$\text{Y}_2\text{O}_3$ system, and by Pang et al. \cite{Pang2019} in $\text{ZrO}_2$-$\text{Ce}\text{O}_2$ ceramics for the tetragonal to monoclinic transformation.\\

For each of the correspondences, there are four variants of martensite, which are defined by the point group relationship of the tetragonal and monoclinic lattices.
For the basis vectors $\vec{a}_i$ of the tetragonal lattice, its point group $\mathcal{P}(\vec{a}_i)$ is the set of rotations that maps the lattice back to itself. 
Thus, $\mathcal{P}(\vec{a}_i) = \{\mat{R} \in \text{SO(3)}: \mat{R} \text{ is a rotation and } \mat{R}\vec{a}_i = \mu_i^j\vec{a}_j \text{ for } \mu_i^j \text{ (a } 3 \times 3 \text{ matrix)} \text{ satisfying det}(\mu_i^j) = \pm1 \}$.
It is reasonable to assume that the point group of the martensite basis $\vec{b}_i$, given by $\mathcal{P}(\vec{b}_i)$, is a subgroup of the point group of the austenite \cite{Kaushik_2003}.
This group-subgroup relationship between the austenite and the martensite gives rise to symmetry-related variants of the martensite.
The relationship between variant $i$ with the transformation stretch tensor $\vec{U}_i$ and variant $j$ with transformation stretch tensor $\mat{U}_j$ is
\begin{equation}
    \label{eq:variants-sym}
    \mat{U}_j = \mat{R} \mat{U}_i \mat{R}^\text{T},
\end{equation}
Where $\mat{R}$ is in the point group of austenite but not in the point group of martensite, i.e., $\mat{R} \in \mathcal{P}(\vec{a_i}) / \mathcal{P}(\vec{b_i})$.
The number of rotations in the point group is called the order of that group.
The number of variants for tetragonal to monoclinic transformation is given by the ratio of the cardinality ($\#$) of the point group of austenite to the cardinality of the point group of martensite. 
The tetragonal crystal has $\# \mathcal{P}(\vec{a_i}) = 8$ and the monoclinic crystal has $\# \mathcal{P}(\vec{b_i}) = 2$. Hence, this results in a total of 4 variants of the martensite.
\\

\begin{figure}[!ht]
  \centering
  \includegraphics[width=1\textwidth]{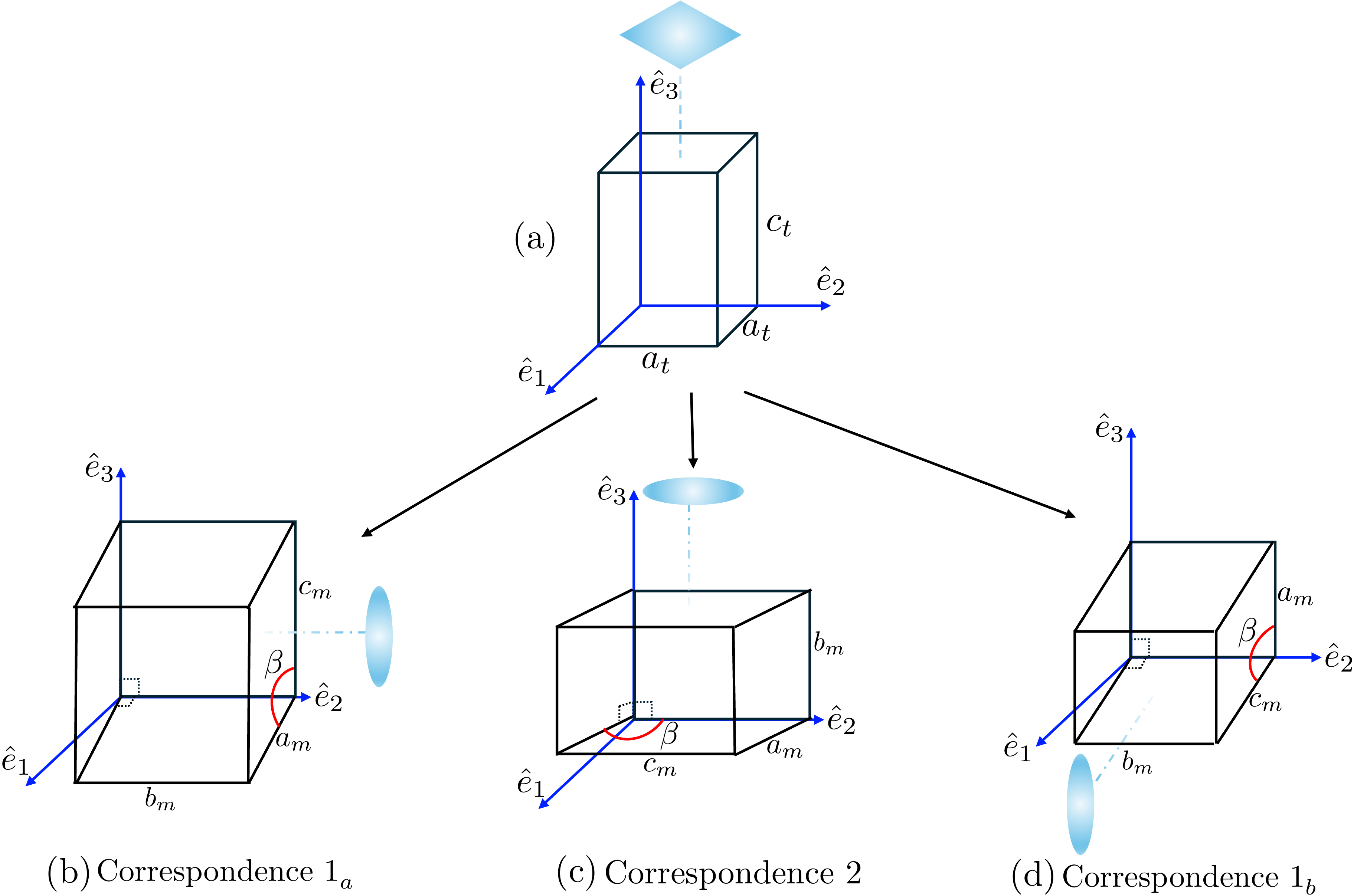}
  \caption[
     plot of correspondences
  ]{The tetragonal crystal in (a) transforms to a monoclinic crystal so that the $c_t$ axis becomes: (b) $c_m$  axis of correspondence $1_a$ (c) $b_m$ axis of correspondence 2 (d) $a_m$ axis of correspondence $1_b$. The gray lens denotes the four fold symmetry axis in tetragonal crystal and two fold symmetry axis in monoclinic crystals.}
  \label{fig_correspondances}
\end{figure}

There are three distinct correspondences ($1_\text{a}$, $1_\text{b}$ and 2) for tetragonal to monoclinic transformation.
The correspondences $1_\text{a}$, $1_\text{b}$ and $2$ are denoted as correspondences C, A and B respectively in Kelly's notation. 
The lattice transformations are categorized into three primary correspondences: Correspondence-$1_\text{a}$ describes the deformation of the tetragonal 4-fold $c_\text{t}$ axis ($[0\ 0\ 1]$) in Fig. \ref{fig_correspondances}a into the monoclinic $c_\text{m}$ axis in Fig. \ref{fig_correspondances}b. In Correspondence-2, the $c_\text{t}$ axis transforms into the monoclinic 2-fold $b_\text{m}$ axis (Fig. \ref{fig_correspondances}c). Finally, Correspondence-$1_\text{b}$ is defined by the mapping of the $c_\text{t}$ axis to the monoclinic $a_\text{m}$ axis (Fig. \ref{fig_correspondances}d).
 \\

Each one of the correspondences has four variants of martensite. For example, the 4 stretch matrices of variants for the correspondence 2 are as follows:
\pagebreak
\begin{equation}
    \label{eq:variants-mech2}
    \mat{U}_1^{(2)} = 
    \scalebox{1}{$
        \begin{bmatrix}
            a & d & 0\\
            d & b & 0\\
            0 & 0 & c
        \end{bmatrix}
    $},\,
    \mat{U}_2^{(2)} =
    \scalebox{1}{$
        \begin{bmatrix}
            b & d & 0\\
            d & a & 0\\
            0 & 0 & c
        \end{bmatrix}
    $},\,
    \mat{U}_3^{(2)} =
    \scalebox{1}{$
        \begin{bmatrix}
            a & -d & 0\\
            -d & b & 0\\
            0 & 0 & c
        \end{bmatrix}
    $},\,
    \mat{U}_4^{(2)} =
    \scalebox{1}{$
        \begin{bmatrix}
            b & -d & 0\\
            -d & a & 0\\
            0 & 0 & c
        \end{bmatrix}
    $}
\end{equation}\\

The unknowns $a$, $b$, $c$ and $d$ are functions of the lattice parameters $a_\text{m}$, $b_\text{m}$,
$c_\text{m}$, $\beta$, $a_\text{t}$ and $c_\text{t}$.
Gu et al. \cite{gu2021exploding} provide the form of stretch matrices and the values for the unknowns $a$, $b$, $c$, and $d$ in Supplementary Section 1. \\

Any variant from one correspondence could form a compatible interface with any variants of the other two correspondences in the twin structure. However, the energy in the transition layer should be made arbitrarily small if they are compatible. 
Gu et al. calculated the number of compatible variants among correspondences for the twin structure and showed them in the tables of supplement \cite{gu2021exploding}.
They observed that a twin structure with variants from mixed correspondences such as $(\mat{U}_i^{(1_\text{a})}/ \mat{U}_j^{(2)}; \text{ }\mat{U}_i^{(1_\text{b})}/ \mat{U}_j^{(2)}; \text{ }\mat{U}_i^{(2)}/ \mat{U}_j^{(2)})$ is favored compared to unmixed correspondences ($\mat{U}_i^{(1_\text{a})}/ \mat{U}_j^{(1_\text{a})}$; $\mat{U}_i^{(1_\text{b})}/ \mat{U}_j^{(1_\text{b})}$; $\mat{U}_i^{(1_\text{a})}/ \mat{U}_j^{(1_\text{b})}$). However, this observation was obtained with data from ceramic system having limited set of dopants. Thus, for any unexplored dopant addition in the ceramic, variants from the unmixed correspondences could also become compatible at the twin interface. \\

\subsection{Identification of shape memory ceramics with low hysteresis} \label{sec: new criteria}

The exact satisfaction of the cofactor condition $q(f) = \det(\mat{C} - \mat{I})$ implies that $q(f)$ vanishes identically if and only if $q(0) = 0$ and $q^{\prime}(0)=0$ (Fig. \ref{fig_meaning_cofactor}c).
Gu et al. evaluated the cofactor conditions for various compositions, noting that while they are never exactly satisfied, they may be considered approximately satisfied at certain compositions \cite{gu2021exploding}.
The approximate satisfaction of the cofactor condition is based on a criteria of minimizing the maximum deviation from the exact satisfaction of the cofactor condition.
The maximum deviation is measured as the maximum value of the quadratic function $|q(f)|$ for values of $f$ between 0 and 1.
The composition with lattice parameters giving the minimum value of the maximum deviation is a potential sample $s$ that could show low hysteresis.
This would physically mean that the free energy of transition layer between the laminate and the austenite can be made small.
Thus, the new criterion based on the approximate satisfaction of the cofactor condition is:
\begin{equation}\label{maxdeviation}
    h_s := \min_{\text{sample } s}\left(\max_{0 \leq f \leq 1} |q^{(s)}(f)| \right)
\end{equation}

This criterion have been applied to find correlation between phase compatibility and efficient energy conversion in Zr-doped Barium Titanate $\text{Ba}(\text{Ti}_{1-x}\text{Zr}_x)\text{O}_3$ having cubic to tetragonal transformation \cite{wegner_correlation_2020}.
The tuning of lattice parameters by changing doping levels of Zr in $\text{Ba}(\text{Ti}_{1-x}\text{Zr}_x)\text{O}_3$ for improved crystallographic compatibility gives significant improvement of transformation and ferroelectric energy conversion properties. 
These lead-free piezoceramics show a close satisfaction of cofactor condition ($h_s = 1.75e^{-8}$) at mole fraction $x = 0.017$ with thermal hysteresis $\mathrm{\Delta T}= 3.93 \, $K. Also, the middle eigenvalue value is found to be very close to 1 with $\lambda_2 = 0.9991$. Similarly, Gu et al. \cite{gu2021exploding} utilized this criteria to find the lowest hysteresis in ceramic $(\text{Zr}_{0.45}\text{Hf}_{0.55}\text{O}_2)_{0.775}-(\text{Y}_{0.5}\text{Nb}_{0.5}\text{O}_2)_{0.225}$ at $y = 0.45$ with thermal hysteresis $\mathrm{\Delta T} = 134$ \degree C. They also observed that at $y=0.45$, an equidistant condition is satisfied, which is as follows:
\begin{equation}
\label{equidistance_condition}
    |\lambda_2^{(1_a, 1_b)}-1|=|\lambda_2^{(2)}-1|
\end{equation}

Where $\lambda_2^{(1_a)}$, $\lambda_2^{(1_b)}$ and $\lambda_2^{(2)}$ are the middle eigenvalues of the deformation stretch tensors of the correspondences 1$_\text{a}$, 1$_\text{b}$ and 2 respectively. We generate synthetic compositions and apply the new criteria mentioned in Eqs. \ref{maxdeviation} and \ref{equidistance_condition} to all compositions to search for SMC. 

%%%%%%%%%%%%%%%%%%%%%%%%%%%%%%%%%%%%%%%%%%%%%%%%%%%%%%%%%%%%%%%%%%%%%%%%%%%%%%%%%%%%%%%%%%%%%%%%%%%%%%%%%%%%
%NEW SECTION%
%%%%%%%%%%%%%%%%%%%%%%%%%%%%%%%%%%%%%%%%%%%%%%%%%%%%%%%%%%%%%%%%%%%%%%%%%%%%%%%%%%%%%%%%%%%%%%%%%%%%%%%%%%%%
\section{Analysis of compositions}

\subsection{Synthetic compositions and prediction of its properties}
The generation of synthetic compositions is essential for identifying an SMC with tuned lattice parameters to meet the new criteria. To achieve this, the mole fractions of each constituent oxide are incremented within the bounds established by the experimental data. Restricting the bounds of mole fraction is crucial for the ML models to accurately predict the lattice parameters of the synthetic composition.
The synthetic dataset was generated through a systematic parametric sweep of the molar fractions. Specifically, the (ZrO$_2$)$_{m_1}$ content was incremented in 0.1\% steps from $m_1 = 0.1675$ to $0.71$. For every $m_1$ interval, the (Y$_{0.5}$Ta$_{0.5}$O$_2$)$_{m_4}$ fraction was similarly varied in 0.1\% increments, followed by a nested 0.5\% stepwise variation of (Er$_2$O$_3$)$_{m_3}$. This iterative approach yielded a comprehensive library of 5,416 distinct synthetic compositions.\\
 
We derive physical features $\vec{x}^*=(\textit{cs},\textit{rio},\textit{en}, \textit{rsla})$ for the synthetic compositions using Eq. \ref{eq:featureX} and predict its transformation temperature $T^*_\text{f}$ using $\vec{x}^*$ as input of the GP model described in Sec. \ref{sssec:num2.2}.
For each $\vec{x}^*$, the GP model predicts a probability distribution $p(T^*_\text{f} \mid \vec{x}^*, (\vec{x}, T_\text{f})) \sim \mathcal{N}(\mu_\text{T}, \sigma_\text{T})$ for $T^*_\text{f}$ instead of a single point estimate. Hence, the predicted $T^*_\text{f}$ has a normal distribution with mean $\mu_\text{T}$ and standard deviation $\sigma_\text{T}$.\\

The GP$_\text{m}$ and GP$_\text{t}$ models for predicting lattice parameters require a 
discrete value of $T_\text{f}^*$ rather than its full probability distribution. 
To account for the uncertainty in $T_\text{f}^*$, we draw $N=15,000$ random samples 
from the predicted distribution $p(T^*_\text{f} \mid \vec{x}^*, (\vec{x}, T_\text{f}))$ 
for each synthetic composition and then paired with its key physical 
features to form the input set:
$\{(T_i, \textit{eff, rcle, cs, rio}) \mid i \in \{1,2,...,N\} \}$. For each input $i$, the GP$_\text{m}$ model outputs a normal distribution $\mathcal{N}(\mu_i, \sigma_i)$. To obtain a single predictive distribution for a synthetic composition, we merge these N individual distributions into a single Gaussian, $\mathcal{N}(\bar{\mu}_m, \bar{\sigma}_m)$, using the following expressions:
\begin{equation}
    \bar{\mu}_m = \frac{1}{N}\sum_{i=1}^{N} \mu_i\\
\end{equation}
\begin{equation}
    \bar{\sigma}_m = \sqrt{\frac{1}{N}\sum_{i=1}^N(\mu_i^2 + \sigma_i^2)-\bar{\mu}_m^2}
\end{equation}
\\

Similarly, samples drawn from the distribution $\mathcal{N}(\mu_{\text{T}}, \sigma^2_{\text{T}})$ of $T_\text{f}^*$ are also combined with $\vec{x^*} = (\textit{eff, cs, en})$ to create the input space of the GP$_\text{t}$ model to predict the probability distribution of tetragonal lattice parameters for each synthetic composition.
This approach of introducing probability distribution of $T_\text{f}^*$ into the GP$_\text{m}$ and GP$_\text{t}$ model allows the propagation of the uncertainty associated with $T_\text{f}^*$--quantified by the 95$\%$ confidence interval  $(\mu_\text{T} - 1.96\sigma_\text{T}, \mu_\text{T} + 1.96\sigma_\text{T})$ of the distribution $\mathcal{N}(\mu_\text{T}, \sigma_\text{T})$--to the uncertainty associated with the distribution of the lattice parameters. \\

The predicted lattice parameters for all synthetic compositions allow for the calculation of cofactor conditions and identify compositions that closely satisfy the cofactor condition.
The lattice parameters are used to calculate the average deformation tensor $\mat{U}_\text{a}$.
For example, for the case of the twinned laminate that has a volume fraction $(1-f)$ of correspondence-$1_\text{b}$ and a volume fraction $f$ of correspondence-2 ($\mat{U}_i^{(1_\text{a})}/ \mat{U}_j^{(2)}$), the average deformation stretch tensor is
$\mat{U}_\text{a} = \mat{U}_i^{(1_\text{b})} + f(\vec{\hat{a}} \otimes \vec{n})$, which is used $q(f) = \det(\mat{U_\text{a}}^\text{T}\mat{U}_\text{a}-\mat{I})$. 
Similarly, $\mat{U}_\text{a} = \mat{U}_i^{(1_\text{a})} + f(\vec{\hat{a}} \otimes \vec{n})$ for the case of twinned laminate having volume fraction $(1-f)$ from the correspondence-1$_\text{a}$ and volume fraction $f$ from the correspondence-2.
To evaluate the criteria mentioned in Eq. \ref{maxdeviation} for each sample, the max$|q(f)|$ value for all mixed correspondence cases $(\mat{U}_i^{(1_\text{a})}/ \mat{U}_j^{(2)}, \mat{U}_i^{(1_\text{b})}/ \mat{U}_j^{(2)}, \mat{U}_i^{(2)}/ \mat{U}_j^{(2)})$ are calculated. Similarly, to evaluate the equidistance condition mentioned in Eq. \ref{equidistance_condition}, the middle eigenvalue $\lambda_2^{(1_\text{a})}, \lambda_2^{(1_\text{b})} \text{ and } \lambda_2^{(2)}$ for each sample are also calculated. \\ 

\subsection{Identification of synthetic composition for experiment}

\begin{figure}[!ht]
  \centering
  \includegraphics[width=1.05\textwidth]{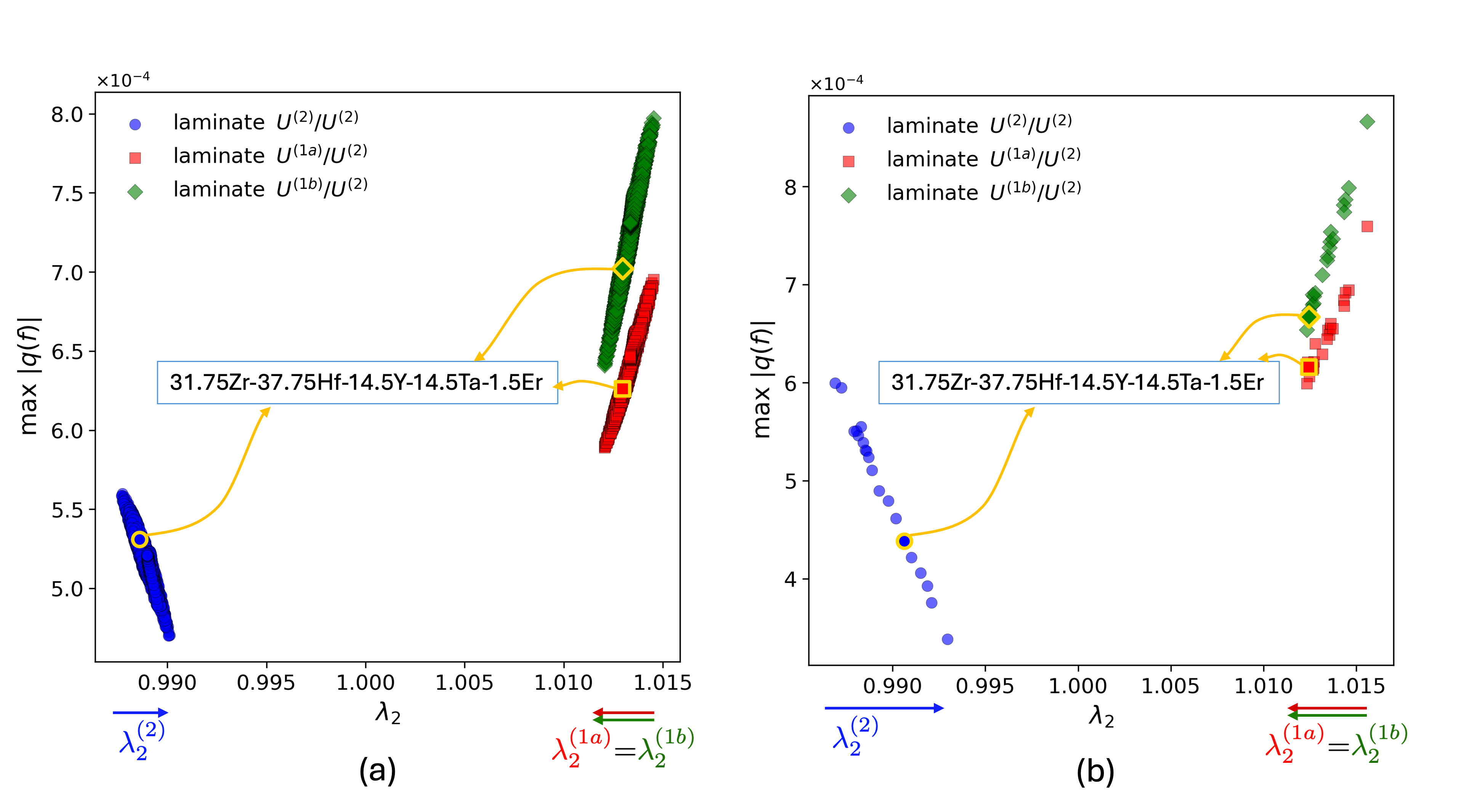}
  \caption[
     Feature selection for tetragonal lattice parameters
  ]{Correlation between $\lambda_2^{(\text{1a})}$, $\lambda_2^{(\text{1b})}$, $\lambda_2^{(\text{2})}$ and max $|q(f)|$ for lattice parameters of: (a) synthetic compositions predicted by the ML model (b) actual compositions measured by XRD-experiment. The highlighted data point in yellow color is selected based on predicted $\lambda_2^{(2)}$ closest to 1 while satisfying the equidistant condition $|\lambda_2^{(1a, 1b)}-1|=|\lambda_2^{(2)}-1|$.}
  
  \label{fig: l2 vs maxqf}
\end{figure}

To search for low-hysteresis SMC within synthetic compositions, the criteria in Eqs. \ref{maxdeviation} and \ref{equidistance_condition} must be closely satisfied. That is, a composition is selected by seeking the minimum value of $|\lambda_2^{(1a, 1b)}-1|-|\lambda_2^{(2)}-1|$,  alongside the minimization of max$|q(f)|$. The predicted lattice parameters $\bar{\mu}$ of the synthetic compositions are used to calculate the stretch tensor $\mat{U}$ of the correspondences $1_\text{a}$, $1_\text{b}$ and $2$ and their middle eigenvalues. The calculated values of middle eigenvalues $\lambda_2^{(1_\text{a})}, \lambda_2^{(1_\text{b})} \text{ and } \lambda_2^{(2)}$ are compared with the corresponding values of $\text{max}|q(f)|$ for twin structures exhibiting mixed correspondences $(\mat{U}_i^{(1_\text{a})}/\mat{U}_i^{(2)})$, $(\mat{U}_i^{(1_\text{b})}/\mat{U}_i^{(2)})$, and $(\mat{U}_i^{(2)}/\mat{U}_i^{(2)})$ respectively. As shown in Fig. \ref{fig: l2 vs maxqf}a, a strong correlation between $\lambda_2$ and $\text{max}|q(f)|$ is observed in all 3 cases. To verify if the same trend is also observed in the compositions fabricated for the XRD-experiments, we randomly chose few compositions and plotted the $\lambda_2$ vs. $\text{max}|q(f)|$ to observe a similar trend as shown in Fig. \ref{fig: l2 vs maxqf}b.\\
 
Further analysis reveals that the strong correlation between $\lambda_2$ and max$|q(f)|$ stem from the fact that most max$|q(f)|$ values are observed at the boundary points $f=0$ and $f=1$ within the domain $0<f<1$. At the boundary points, the twin structure having multiple variants of martensite becomes a single variant. To better understand this, lets take a special case of twin laminate having correspondences $\mat{U}_i^{(1_\text{b})}/\mat{U}_j^{(2)}$. If the maximum deviation exists at $f=0$, $q(f)$ simplifies to:
\begin{equation}
\label{simplifiedcofactorconditions}
q(f)= \det[(\mat{U}_i^{(1_\text{b})})^\text{T}\mat{U}_i^{(1_\text{b})} - \mat{I}] = ((\lambda_1^{(1_\text{b})})^2-1)((\lambda_2^{(1_\text{b})})^2-1)((\lambda_3^{(1_\text{b})})^2-1) = 0     
\end{equation}
Similarly, if the maximum deviation exists at $f=1$, $q(f)$ simplifies to:
\begin{equation}
q(f)= \det[(\mat{U}_i^{(2)})^\text{T}\mat{U}_i^{(2)} - \mat{I}] = ((\lambda_1^{(2)})^2-1)((\lambda_2^{(2)})^2-1)((\lambda_3^{(2)})^2-1) = 0
\end{equation}
Alternatively, the condition $\lambda_2^{(1_\text{b})} =1$ at $f=0$,  and $\lambda_2^{(2)}=1$ at $f=1$ serves as simplified alternatives to the maximum deviation criterion for twin system with correspondences-1$_\text{b}$ and 2.
Thus, for the set of synthetic compositions $C$, the max$|q(f)|$ criteria can be alternatively stated for finding single, specific composition $c$ as:
\begin{equation}\label{alternativemaxqf}
    \min_{c \in C} \left( \max \left\{ |\lambda_2^{(1_\text{a}, 1_\text{b})} - 1|, |\lambda_2^{(2)} - 1| \right\} \right)
\end{equation}

By selecting the composition that satisfy the criterion in Eq. \ref{alternativemaxqf}, we naturally prioritize configurations where the larger of the two deviations is suppressed. In this minimax strategy, the global minimum of this objective function is reached when the two deviations are balanced, i.e., $|\lambda_2^{(1a, 1b)}-1|-|\lambda_2^{(2)}-1|$. Compositions deviating from this equality are inherently limited by whichever eigenvalue's deviation is larger, rendering them suboptimal compared to the balanced state. This is also the experimentally observed equidistant condition observed by Gu et al. \cite{gu2021exploding}.\\

\subsection{Experimental measurements \& comparison with predictions}
The thermal measurements of transformation temperature and transformation enthalpy are performed by using a TA instruments SDT 650 machine capable of doing DTA measurements with a temperature range between room-temperature and 1500 °C. For samples that have one or more of the four temperatures used for thermal characterization below RT, a Netzsch DSC 204 F1 Phoenix with a minimum temperature of -160 °C can be used. With the combined temperature range from -160 °C to 1500 °C a throughout measurement of $M_{s}$, $M_{f}$, $A_{s}$ and $A_{f}$ for a large range of sample composition can be ensured.\\

For the determination of lattice parameters all samples underwent temperature controlled XRD investigations. The Rigaku SmartLab 9kW was employed for this task and all measurements where performed using mainly Cu $K_{alpha}$ radiation. A Ni filter was used to reduce the intensity of Cu $K_{beta}$ wavelength without the need for proper monochomatization. For temperature control the Anton Paar temperature stages DHS 1100 and DCS 350 where used. Those two stages combined allowed for precise XRD $\theta{}/2\theta{}$ measurements in a temperature range between -100 °C and 1100 °C. Having measured the transformation temperatures for the sample in question beforehand allows for very small temperature intervals during martensitic and austenitic transformations. These measurements moreover allowed to measure not only thermal expansion of individual lattice parameters of both, high and low temperature phase, but by performing Rietveld refinements on each measurement, the determination of relative phase fractions. Since the refinement is more robust with higher signal intensities, we opted to exclude data calculated from minority phases if the phase fraction dropped below 20\%.\\

Those measurements are also used to ensure the absence of any secondary phases which can possibly be formed by limited solubility in the solid solution.
The refinements were performed semi-automatically by using the batch processing capabilities of the software package TOPAS v6. \\

We identified 31.75ZrO$_2$--37.75HfO$_2$--14.5Y$_{0.5}$Ta$_{0.5}$O$_2$--1.5Er$_2$O$_3$ among all synthetic compositions for experiments, since it closely satisfies the equidistant condition $|\lambda_2^{(1a, 1b)}-1|-|\lambda_2^{(2)}-1| = 0.0263$. This composition is also highlighted in Fig. \ref{fig: l2 vs maxqf}.
For the selected composition, we have measured start and finish temperatures of Austenite $A_s = 611 \degree \text{C}$, $A_f = 661 \degree \text{C}$ and Martensite $M_s = 518 \degree \text{C}$ and $M_f = 480 \degree \text{C}$ and calculated the transformation temperature $T^*_f = 0.5*(M_s + A_f) = 589.5 \degree \text{C}$.
We also measured lattice parameters for this sample, as shown in Table \ref{table4:MLvsExp}.
The ML predictions of lattice parameters for this composition are found to be in good agreement with experiments, as shown in Table \ref{table4:MLvsExp}.
From the lattice parameters, we found the actual value of middle eigenvalue of correspondence-2 is $\lambda_2 = 0.9921$, and $\Delta \text{V/V} = 3.65 \%$. 
The thermal hysteresis for this sample is $\Delta T = 137 \degree \text{C}$. 
Although this sample satisfies all design criteria proposed by Pang et al. \cite{Pang2022}-specifically, $\lambda_2^{(2)}=0.9921$, $\Delta \text{V/V} = 3.65 \% $, all dopants are within solubility limits, and $T^*_f = 636.78 \degree \text{C}$—it nonetheless exhibits significantly high hysteresis.\\ 

To understand the role of dopant Er$_2$O$_3$ in reducing the tetragonality ratio $(c_\text{t}/a_\text{t})$, we calculate the $(c_\text{t}/a_\text{t})=1.0258$ for this composition using predicted lattice parameters, and after addition of $1.5 \%$ Er$_2$O$_3$ in this composition, the $(c_\text{t}/a_\text{t})$ ratio reduces to 1.0246, a $0.117\%$ reduction in the tetragonality value.
This implies that small amounts of Er$_2$O$_3$ addition (up to solubility limits) in the Zr-Hf-Y-Ta ceramic system reduces its tetragonality ratio; however, it has a weak effect in reducing tetragonality. 

\begin{table}[h!]
\centering
\caption{Comparison of predicted vs experimental values}
\begin{tabular}{|c|c|c|c|}
\hline
\strut & \parbox{3 cm}{ML Predicted} & \parbox{3cm}{Experimental} & \parbox{3cm}{\% Error}\\
\hline
$T^*_f$ & 638.76 \( \pm \) 65.97 & 589.5 & 8.36 \% \\ 
\hline
$a_m$ & 5.1765 \( \pm \) 0.0021 & 5.1778 & 0.025 \% \\
\hline
$b_m$ & 5.1967 \( \pm \) 0.0067 & 5.1908 & 0.114 \% \\
\hline
$c_m$ & 5.3648 \( \pm \) 0.0070 & 5.3667 & 0.035 \% \\
\hline
$\beta$ & 98.7273 \( \pm \) 0.0971 & 98.7377 & 0.010 \% \\
\hline
$a_t$ & 3.6256 \( \pm \) 0.0075 & 3.6255 & 0.003 \% \\
\hline
$c_t$ & 5.2539 \( \pm \) 0.0186 & 5.2321 & 0.042 \% \\
\hline
\end{tabular}
\label{table4:MLvsExp}
\end{table}

\section{Conclusion} 
In summary, we have presented a Gaussian process (GP) framework capable of accurately predicting both the transformation temperatures and the lattice parameters for the monoclinic and tetragonal structures of a chosen ceramic system. 
By predicting the lattice parameters across a wide range of synthetic compositions, this framework reveals a strong correlation between the middle eigenvalues of the stretch tensor and the maximum deviation from the exact satisfaction of the cofactor conditions. Thus, framework helps in revealing that max$|q(f)|$ criteria is equivalent to the experimentally observed equidistance condition.\\

We selected and fabricated a synthetic composition—31.75ZrO$_2$--\allowbreak 37.75HfO$_2$--\allowbreak 14.5Y$_{0.5}$Ta$_{0.5}$O$_2$--\allowbreak 1.5Er$_2$O$_3$—that closely satisfies the equidistant criterion. Differential thermal analysis (DTA) and X-ray diffraction (XRD) were employed to measure its transformation temperature and lattice parameters, respectively; the experimental results were found to be in good agreement with the values predicted by the GP models. The steps described in the GP framework could also be applied to new ceramic systems having additional dopants for accurate prediction of their transformation temperature and lattice parameters.
Such machine learning models have applications in detecting synthetic compositions with predicted desired properties, thus reducing efforts in experiments by informing targeted values in the compositional space.\\

The selected composition satisfied all four design criteria ($\lambda_2^{(2)}=1$, low volume change, solid solubility, high transformation temperature ($M_\text{s}$ > 500 $\degree \text{C}$))  required to minimize thermal hysteresis in ZrO$_2$-based ceramics.
However, in the differential thermal analysis experiment, we found a high value of thermal hysteresis: $\Delta T = 137 \degree \text{C}$.
Thus, the current four criteria are not universal for ceramics, and they do not always result in low hysteresis ceramic. 
In the future work, there is a need to explore a new dopant that could reduce the tetragonality ratio of the parent phase to an extent that the ceramic system fully transforms from cubic to monoclinic.
This is because the cubic to monoclinic phase transformation results in a greater number of variants of the martensite compared to the tetragonal to monoclinic transformation. 
This allows more variants to orient themselves to allow for large deformation during phase transformation and satisfy the cofactor conditions.\\

\newpage
\bibliography{bibcsm}
\bibliographystyle{unsrt}
\end{document}